\documentclass[12pt]{article}

\usepackage{a4wide,graphicx,cite}

\renewcommand{\baselinestretch}{1.5}
\usepackage[dvips]{color}

\begin{document}

\begin{center}

  {\large \bf Novel structural features of CDK inhibition revealed by an ab initio
                     computational method combined with dynamic simulations}

  \vspace{1cm}
   Lucy Heady,$^1$ Marivi Fernandez-Serra,$^2$
   Ricardo L. Mancera,$^3$ Sian Joyce,$^{1,4}$  \\
   Ashok R. Venkitaraman,$^{5}$
   Emilio Artacho,$^6$ Chris-Kriton Skylaris,$^7$ \\
   Lucio Colombi Ciacchi,$^{1,8,*}$
   and Mike C. Payne$^1$

  \vspace{0.5cm}
{\it
     $^1$Theory of Condensed Matter Group, Cavendish Laboratory, \\
         University of Cambridge, J J Thomson Avenue, Cambridge CB3 0HE, UK \\
     $^2$D\'epartment de Physique des Mat\'eriaux, \\
         Universit\'e Claude Bernard Lyon, 69622 Villeurbanne, France \\
     $^3$Western Australian Biomedical Research Institute \& School of Pharmacy \\
         and School of Biomedical Sciences,  Curtin University of Technology, \\
         GPO Box U1987, Perth WA 6845, Australia  \\
     $^4$Photonics Theory Group, Tyndall National Institute \\
         Lee Maltings, Cork, Ireland \\
     $^5$ CR UK Department of Oncology \& The Medical Research
          Council Cancer Cell Unit, \\
         Hutchison/MRC Research Centre, Hills Road, Cambridge, CB2 2XZ, UK \\
     $^6$Earth Sciences Department, University of Cambridge, \\
         Cambridge CB2 3EQ, UK \\
     $^7$Physical and Theoretical Chemistry Laboratory,\\
         University of Oxford, South Parks Road, OX1 3QZ, UK \\
     $^8$Fraunhofer Institut f\"ur Werkstoffmechanik,   \\
         W\"ohlerstrasse 11, 79108 Freiburg, Germany \\
  }

\end{center}
\vfill
$^*$ Email: lucio@izbs.uni-karlsruhe.de, Fax: +49 761 5142404, Tel.: +49 761 5142113

\newpage

 \begin{center}
    {\bf Abstract}
 \end{center}
The rational development of specific inhibitors for the $\sim$500 protein
kinases encoded in the human genome is impeded by poor understanding
of the structural basis for the activity and selectivity of small
molecules that compete for ATP binding.
Combining classical dynamic simulations with a novel {\em ab initio}
computational approach linear-scalable to molecular interactions
involving thousands of atoms, we have investigated the binding of five
distinct inhibitors to the cyclin-dependent kinase CDK2.
We report here that polarization and dynamic hydrogen bonding effects
--so far undetected by crystallography-- affect both their activity and
selectivity.
The effects arise from the specific solvation patterns of water
molecules in the ATP-binding pocket or the intermittent formation of
hydrogen bonds during the dynamics of CDK-inhibitor interactions, and
explain the unexpectedly high potency of certain inhibitors like
3-(3H-Imidazol-4-ylmethylene)-5-methoxy-1,3-dihydro-indol-2-one (SU9516).
The Lys89 residue in the ATP-binding pocket of CDK2 is observed to
form temporary hydrogen bonds with the three most potent inhibitors.
This residue is replaced in CDK4 by Thr89, whose shorter side-chain
cannot form similar bonds, explaining the relative selectivity of the
inhibitors for CDK2.
Our results provide a generally applicable computational method for
the analysis of biomolecular structures, and reveal hitherto
unrecognized features of the interaction between protein kinases and
their inhibitors.

\newpage

\section{Introduction}

Approximately 500 different protein kinases are encoded in the human
genome~\cite{Manning_2002}.
The similarity of the mechanism and structure of their catalytic
domains remains a major obstacle to the rational development of
specific inhibitors for the treatment of human diseases ranging from
cancer to auto-immunity.
One important case in point involves the family of cyclin-dependent
kinases (CDKs), members of which are essential for progression through
different stages of the cell division cycle in all
eukaryotes~\cite{Morgan_95}.
In particular, one family member, CDK2, is required for the events
that lead to DNA replication~\cite{Loog_05,Brown_99,Russo_96}.
CDK2 and its associated cyclins are over-expressed in cancer cells,
and might contribute to their deregulated growth~\cite{Vermeulen_03}.
Therefore, inhibition of CDK2 through insertion of small molecules
into its ATP-binding pocket has long been a potential target for
cancer therapies~\cite{Vermeulen_03,Huwe_03,Wadler_01,Noble_04}.
A number of inhibitor molecules have already been designed for this
purpose, some of which are currently in clinical trials (see
e.g. Refs.~\cite{Huwe_03,Noble_04} and references therein).
The clinical efficacy of these inhibitors critically depends not only
on their potency --that is, their ability to bind to CDK2 more stably
than ATP-- but also on their selectivity for CDK2 over other, highly
homologous members of the CDK family.
Non-selective inhibition carries the risk of undesired and potentially
toxic side-effects~\cite{Bain_03}.
Thus, along with potency, selectivity is a crucial issue in inhibitor
design~\cite{Noble_04,Park_04,Ikuta_01}.
However, although the crystallographically determined structures of
inhibitors bound to CDK2 have been used as a basis for the rational
development of several different inhibitors, features of their potency
in CDK2 inhibition, or their selectivity for CDK2 over the closely
related kinase CDK4, remain to be explained.

We address here the problem of computing relative potencies and
explaining the selectivity of five CDK2 inhibitors using an approach
that combines classical dynamic simulation with novel {\em ab initio}
calculations linear-scalable to molecular interactions involving
thousands of atoms~\cite{Soler_02,Skylaris_05}.
First principles quantum techniques enable us to calculate binding
energies of hydrogen-bonded systems to high accuracy, while classical 
dynamical simulations allow access to large regions of the potential 
energy surface and long time scales.
As a reference, we also perform a series of extensive docking
simulations and scoring functions binding energy calculations.

Our study is based on the crystallographically determined structures
of inhibitors bound to CDK2.
These provide a clear picture of the binding features between the
ligand and those residues that constitute the binding pocket,
revealing the dominant interactions to be hydrogen bonding,
electrostatic and van der Waals forces.
Potent lead compounds for new CDK inhibitors have been readily
developed by taking advantage of these interactions via analysis of
the so-called structure-activity relationship~%
\cite{Ikuta_01,Hardcastle_04,Gibson_02,Sayle_03,Legraverend_00}.
In addition, computer-aided approaches such as
database~\cite{Gray_98}, docking~\cite{Otyepka_00,Gabb_97,Ducrot_00}
and scoring function methods~\cite{Mancera_04} provide an effective
way of probing the binding modes and testing the binding strength of
large arrays of molecules, and thus help identify potential new lead
compounds.
Moreover, molecular dynamics simulations have proven very useful for
analysis of the binding modes and the structural rearrangements which
are connected with the inhibition or activation of CDKs~%
\cite{Kriz_04,Bartova_04,Otyepka_02,Cavalli_01}.
In this work, we extend these techniques to encompass the use of first
principles methods, which permit calculation of hydrogen-bond strengths 
at an accuracy of about 1~kcal/mol~\cite{Ireta_04}, and do not depend 
on any external set of empirical parameters.

A number of structural studies have demonstrated that the measured
potency of inhibition can be directly correlated with the strength of
the local protein/inhibitor interactions within the ATP binding pocket
(see e.g. \cite{Noble_04} and references therein).
This idea has lead to the rational development of inhibitors via
addition of functional groups designed to encourage specific
interactions, such as hydrogen bonds, with promising residues.
Most inhibitors bind to CDK2 in a fashion similar to the adenine ring
of ATP, forming a triplet of hydrogen bonds to the peptide backbone of
residues Glu81 and Leu83 which reside in the hydrophilic hinge region
at the back of the binding pocket~\cite{Arris_00}.
With the goal of increasing the protein-inhibitor interactions,
specific functional groups have been added to $O^6$-cyclohexylmethylguanine
(NU2058)~\cite{Arris_00} to develop its more potent variant 
$O^6$-cyclohexylmethoxy-2-(4'-sulfamoylanilino)purine (NU6102)~\cite{Davies_02}, 
and to  2,6-diamino-4-cyclohexylmethyloxy-5-nitrosopyrimidine (NU6027)~%
\cite{Arris_00} to obtain a potent carboxamide derivative (the 9d variant 
in~\cite{Sayle_03}) (see Figure~1).
Structural investigations demonstrated that the higher potency of
these two variants is due to formation of additional interactions with
the polar residue Asp86, while the ``standard'' triplet of hydrogen
bonds is retained~\cite{Davies_02,Sayle_03}.
A puzzling exception is 
3-(3H-Imidazol-4-ylmethylene)-5-methoxy-1,3-dihydro-indol-2-one (SU9516) 
which appears to form only the standard hydrogen bond triplet but is 
highly potent and also presents a good degree of selectivity for CDK2 
against CDK4~\cite{Lane_01,Moshinsky_03}.
Unlike potency, the issue of selectivity has proven to be much harder
to explain via analysis of the available crystal structures
only~\cite{Park_04,Ikuta_01}.
In the particular case of SU9516, the observed selectivity had
previously been proposed to arise from an interaction with the CDK2
specific residue Lys89~\cite{Lane_01}, but no such interaction was
found in a recent crystallographic study~\cite{Moshinsky_03}.
Indeed, the Lys89 residue was also a target for the two derivatives
mentioned above, NU6102 and 9d-NU6027, but again no such interaction
has been observed in the crystallographic studies
performed~\cite{Sayle_03,Davies_02}.

The potency of inhibition can be quantitatively described by the
so-called IC$_{50}$ value, i.e. the concentration of inhibitor which
is required to reduce the activity of the protein by 50~\% with
respect to a chosen reference state in the absence of inhibitor.
We note that the IC$_{50}$ value is dependent on the specific
experimental conditions, in particular on the ATP concentration
used in the activity assay.
A more objective measure of the potency, which can be used to compare
results from different assays, is the inhibition constant $K_i$.
This is defined as the equilibrium dissociation constant for the
reaction of an inhibitor with a protein to form an inhibitor-protein
complex.
Under the same assay conditions, $K_i$ and IC$_{50}$ values are
related via a simple proportionality relationship~\cite{Cheng_73},
so that often the values can be directly compared~\cite{Sayle_03}.
Therefore, both IC$_{50}$ and $K_i$ values can be assumed to be
directly related to the free energy change associated with the
inhibitor binding to the protein~\cite{Davies_02}.
Our ultimate aim here is to reproduce the experimentally measured relative
potencies of the five inhibitors mentioned above against CDK2 by
calculation of the differences between their free energies of binding.

Our calculations reveal that a static analysis of the hydrogen bond
patterns formed by the inhibitors in the ATP binding pocket is
generally not sufficient to predict the correct rank order of 
inhibitor potencies.
In particular, electronic structure calculations at the DFT level
appear to be necessary to account quantitatively for the screening
of the hydrogen bond interactions due to the presence of water molecules
in the binding cleft.
Moreover, dynamical simulations reveal the profound effect of the
motion of certain key residues that are involved in hydrogen bonding
within the binding pocket.
Only by taking into account the dynamical nature of the protein/ligand
interactions are we able to rationalise the measured potency and
explain the observed CDK2 selectivity of the inhibitors considered.

\section{Free energy of binding and potency of inhibition}

The potency of an inhibitor is measured in terms of the inhibition
constant $K_i$, which is inversely proportional to the association
constant $K_a$ of the reaction
\begin{equation}
 \mbox{Protein\ }(P) + \mbox{Ligand\ } (L) \rightarrow
 \mbox{Protein-Ligand Complex\ } (PL).
\end{equation}
This process involves a change of free energy $\Delta G =
-k_BT\ln K_a$, where $k_B$ is the Boltzmann constant and $T$ is the
absolute temperature.
Let us first split $\Delta G$ into gas-phase and solvation contributions,
and the former in enthalpic and entropic contributions:
\begin{equation}
 \Delta G = \Delta G_g + \Delta G_{solv} = 
 \Delta E_g - T\Delta S_g + \Delta G_{solv} \,,
\end{equation}
where $\Delta  G_{solv}$ is defined as the difference 
$ G_{solv}(PL) -  G_{solv}(P) - G_{solv}(L)$, and 
$\Delta  E_g$ is defined as $ E_g(PL) - E_g(P) - E_g(L)$.
The enthalpic term $\Delta E_g$  can be accurately calculated using the DFT
approach, whereas both the solvation free energy contributions $\Delta G_{solv}$ 
and the gas-phase entropic contributions $\Delta S_g$ are in general not readily
accessible within this computational scheme.
However, when we are looking at the relative potencies between
two inhibitors 1 and 2 then we are only interested in the ratio
${K_i^1}/{K_i^2}$. 
This enables us to look at the difference between their
associated free energy gains, $\Delta\Delta G =
\Delta G^1 - \Delta G^2$, which can be directly compared with the $K_i$ values
via the relationship:
\begin{equation}
 \Delta G^1 - \Delta G^2 = k_B T \ln\frac{K_i^1}{K_i^2}\;,
 \label{eq:ddg2}
\end{equation}
and can be written as follows:
\begin{equation}
 \Delta\Delta G = \Delta\Delta E_g +  \Delta\Delta G_{solv}  - T \Delta\Delta S_g.
\label{eq:deltag}
\end{equation}
Following Ref.~\cite{Wang_01}, we can assume that the difference between 
the gas-phase entropy changes $\Delta\Delta S_g$ is negligible within the 
error bar associated with the calculation of the remaining terms.
This appears to be justified in the present case given 
the large similarity between the systems considered.

Within this approximation, we are left with only two terms: 
\begin{equation}
\Delta\Delta G = \Delta\Delta E_g + \Delta\Delta G_{solv}.
\end{equation}
We calculate the gas phase enthalpy differences $\Delta E_g$ within the 
DFT approach, and the solvation free energy differences $\Delta G_{solv}$ 
classically, using the Generalised Born/Surface Aerea (GBSA) model.
Continuum theories have been used to compute free energies of solvation with
success for over 20 years~\cite{Cramer_99,Jayaram_98,Sitkoff_94},
accounting for entropic and hydrophobic effects in a simple and
computationally inexpensive manner.
This allows us to avoid all the difficulties involved in first principles 
simulations of polar solvents~\cite{Galli_04,Fernandez_04,Vondele_05}.
On the other hand, we are aware that more refined formalisms could be 
used to treat solvation effects to higher accuracy~\cite{Mancera_wat}, 
and in particular we will pay great attention to the specific hydration 
patterns of the $PL$ complexes within the binding pocket, including 
explicit solvent water molecules in our final set of calculations 
(Section 4.1).
Finally, using classical force-field techniques we have checked that 
dispersion forces, which are severely underestimated in the standard 
DFT approach, contribute in a nearly equal way to the binding energy of
all inhibitors to CDK2, and therefore do not contribute to the free 
energy differences $\Delta\Delta G$ (see Computational Details).

\section{Static binding energy calculations}

\subsection{Convergence of binding energy with the system size}

Within the approximations described in the previous section, we are in
principle able to compute the relative differences of binding free energy
$\Delta\Delta G$, provided that we succeed in calculating the total
energy (enthalpy) of the protein and the protein/ligand complexes.
Inclusion of the whole CDK protein in our calculation would be
prohibitively expensive in terms of computer time, but this is
unnecessary in the present case.
Since the small inhibitors bind to a well defined and spatially
limited region of the protein (the ATP binding pocket), we are able to
consider only those amino acid residues which are close enough to the
inhibitor to significantly contribute to the binding energy.
In doing this we must strike a delicate balance between the gain in
accuracy from using a DFT approach and the loss of a realistic model due
to limiting the size of the systems considered.
In order to select the smallest possible fragment of CDK2 which 
ensures that all important interactions in the pocket are taken
into account we tested the convergence of $\Delta E_g$ with respect 
to increasing fragment size.

The inhibitor used in this convergence test is staurosporine
(inhibitor {\bf 7} in Figure~1).
Due to its large size, the calculated binding energy of
staurosporine to CDK2 is expected to be strongly influenced by
long-range electrostatic interactions, and hence to be very sensitive
to the size of the system used to model the binding pocket.
Our starting point is the available crystal structure of staurosporine
in complex with CDK2 (PDB key 1AQ1)~\cite{Lawrie_97}.
As a first model, we consider staurosporine surrounded by only those
amino acids with which it makes direct hydrogen bonds (E81, L83, D86
and Q131).
Since the side chains of E81 and L83 do not directly interact with
the inhibitor, they are replaced by methyl groups bound to the
respective C$_{\alpha}$ (Figure~2).
Keeping the C$_{\alpha}$ atoms of the protein backbone constrained in
their original x-ray positions, we fully minimise the geometry of the
system using the DFT approach (see Computational Details).
This results, in particular, in the accurate estimation of the
hydrogen bond lengths and angles between the inhibitor and the
protein.
Three larger models are then constructed, adding to the minimised
small system all amino acids within 5.0~\AA, 7.0~\AA, and 10.0~\AA\
from the inhibitor (Figure~2).
The binding energy is then calculated for all models as the difference
in total energy between the protein/inhibitor complex, the isolated
protein model, and the isolated inhibitor.
With the exception of the smallest model, we choose not to minimise
the geometry of the systems, for the following reasons.
Firstly, the potential energy surface of a large number of amino acids is
expected to be very flat and contain a large number of local minima
so that (i) the minimisation would take a long time and would be very
expensive to perform at the quantum level of accuracy, and (ii) even
after full minimisation it is very hard to know whether the correct global
minimum has been found.
Secondly, in principle we should minimise both the protein/inhibitor
and the isolated protein models.
However, in this case the binding energy would be affected by differences
in the relative positions of amino acids far away from the inhibitor,
with the risk of wrongly estimating the binding energy because of the
aforementioned difficulties in finding the global energy minimum.
Therefore, we assume here that the available crystal structures
represent an average configuration of the amino acids reliable enough
to compute long range interactions to an accuracy within the
error bars associated with all other approximations.
Binding energies of systems which are fully relaxed using the
classical force field approach will be presented in Section~5.

The calculated binding energy values using both the \textsc{Siesta}
and the \textsc{Onetep} codes are reported in Figure~2 alongside the 
model systems used.
We find that the binding energy converges to within 0.15~kcal/mol for
the third model of the binding pocket, i.e. when all amino acids at a
distance of 7~\AA\ from the inhibitor are explicitly considered.
In the next section, models of this size will be used to compute the
binding energy of all other inhibitors considered using the O(N) DFT
approach (see Computational Details).
As a further validation test of the O(N) DFT techniques employed, we
have calculated binding energy values for the first model of Figure~2
using also the traditional plane-wave formalism.
The computed binding energies obtained with the \textsc{Castep},
\textsc{Siesta} and \textsc{Onetep} codes are -32.0~kcal/mol,
-32.2~kcal/mol and -32.4~kcal/mol, respectively.
This confirms that both O(N) DFT approaches faithfully reproduce with 
equivalent accuracy the results of the traditional PW implementation.

\subsection{Static calculations using the available cystal structures}

As mentioned in the Introduction, we consider in our study the
inhibitors NU2058 ({\bf 1}), NU6027 ({\bf 2}), NU6102 ({\bf 3}),
the 9d-variant of NU6027 ({\bf 4}), and SU9516 ({\bf 5}) (Figure~1)
(for simplicity, inhibitors are referred to by their associated number
in the following text).
Experimentally determined values of $K_i$ are available for the
inhibitors {\bf 1}, {\bf 2}~\cite{Arris_00}, {\bf 3}~\cite{Davies_02},
and {\bf 5}~\cite{Moshinsky_03}.
Moreover, IC$_{50}$ values for {\bf 1}, {\bf 2}, {\bf 3}, and {\bf 4}
have been determined in the same assay~\cite{Sayle_03}.
Therefore we are able to obtain an estimate of the $K_i$ value for
{\bf 4} using the proportionality relationship between IC$_{50}$ and
$K_i$~\cite{Cheng_73}.
The experimental K$_{i}$ values of all these inhibitors are reported
in Table~1.

To investigate the interactions between the inhibitors and the
protein, we start from the crystallographic determined structures of
the $PL$ complexes which are available in the RCSB Protein Data Bank.
Here one has a choice between structures where the inhibitors are bound
to the {\em active} form of CDK2, i.e. the Thr160-phosphorylated CDK2/cyclin A
complex, or structures where they are bound to the {\em inactive}  form,
i.e. monomeric CDK2.
A major structural difference between the two forms lies in the
availability (in the inactive form) vs. unavailability (in the active
form) of the K33 residue within the binding pocket.
It has been suggested that, when available, K33 strongly interacts with
the natural ATP ligand in a way which suppresses its turnover~%
\cite{Brown_99,Russo_96}.
Therefore, the binding energy of inhibitors to the inactive form of CDK2
is expected to be affected by the presence of K33 in the binding pocket.
Namely, the presence of the K33 ligand in the binding pocket
would result in spurious interactions which are in fact absent in the
actual activity assays from which the $K_i$ values are extracted.
Therefore, since our aim is to present a consistent comparison of the
binding energies of the inhibitors related to their inhibition activity,
we choose to consider only the interactions between the inhibitors and
the binding pocket of CDK when K33 is not available.
Indeed, the active CDK2/cyclin A complex is increasingly used as the
reference structure in structure-activity-relationship investigations
of drug activity~\cite{Davies_02}.

In the RCSB database, structures are available for {\bf 1}, {\bf 3}
and {\bf 4} bound to the Thr160-phosphorylated cyclin-A/CDK2
complex (PDB keys are 1H1P, 1H1S, 1OGU, respectively), and for
{\bf 2} and {\bf 5} bound to the monomeric CDK2 form (PDB keys
are 1E1X and 1PF8, respectively).
In order to prevent the calculated binding energies from being affected by
strong spurious interactions with the polar residue K33, this has
been substituted with an alanine residue in the two latter cases.
In the available crystal structures, all inhibitors are observed to
form a triplet of hydrogen bonds with E81 and L83.
In addition, D86 is observed to form bonds with {\bf 3} and
{\bf 4}, which leads to their increased potency.

Calculations of the binding energies of all inhibitors are performed
as described in the case of staurosporine (Section~3.1), considering
all amino acids within 7~\AA\ from each inhibitor.
Namely, the hydrogen bond distances are optimised using the
PW-DFT approach considering only those amino acids that
are directly bound to the inhibitors in the crystal structure.
Then this optimised model is embedded in the larger pocket model, and
the binding energy $\Delta E_g$ is calculated using the O(N) DFT
approach without any further geometry optimisation.
The obtained binding energy values are reported in Table~1 along with
the values of the solvation free energy differences $\Delta G_{solv}$
computed within the GBSA model.
From these values we are able to compute the relative free energy
differences between the inhibitors, $\Delta\Delta G$, using
equation~5, where we take as a reference the binding energy value 
of {\bf 1}, which is the least potent inhibitor considered.
As a comparison, we report in Table~2 the values of
binding energy differences calculated with three different 
sets of scoring functions after extended docking simulations,
as described in the Computational Details section.

The results of the docking simulations appear to be roughly consistent
with the measured rank of potencies (except in the case of 
inhibitor {\bf 2}).
However, quantitative agreement between the measured and calculated
binding energies could not be achieved, the values of $\Delta\Delta G$
being consistently underestimated.
On the other hand, the agreement between the relative energies of
binding calculated from static structures within the DFT formalism and
the experimentally determined relative potencies is worse.
In particular, the binding energies of inhibitor {\bf 3} is highly overestimated 
with respect to the binding energy of inhibitor {\bf 1}.
Furthermore, the wrong rank order of potency is predicted for
inhibitor {\bf 2} and, more dramatically, for {\bf 5}.
We note that the binding energy of {\bf 5} is far too small even
when only the direct hydrogen bond interactions are considered.
In this case a strong interaction (e.g. a direct hydrogen bond) 
seems to be missing in the model considered.
However, the recently resolved x-ray structure of this inhibitor bound
to CDK2~\cite{Moshinsky_03}, from which our model was constructed, reveals
no interactions other than the usual triplet of hydrogen bonds with
the backbone of E81 and L83.

In general, the inconsistencies between the measured and calculated
values are too large to be due to approximations such as the
neglection of gas-phase entropic contributions, or the limited 
size of the models considered.
Rather, these results suggest that the precise solvation patterns of the 
$PL$ complexes need to be explicitly considered, and a more careful
analysis of the potential energy surface of the binding modes needs to
be performed.
In the next section we tackle these issues in detail by performing a
dynamical analysis of the binding modes of the inhibitors bound to the
active form of CDK2 in the explicit presence of water solvent
molecules.

\section{Dynamical force field simulations}

Our investigations start with a dynamical analysis of the isolated CDK2 protein
in explicit water solvent (PDB key 1HCL) and bound to its
natural ligand, ATP (PDB key 1HCK)~\cite{Schulze_96}.
After minimisation and equilibration as described in the Computational
Details section, we perform a 0.5~ns run for each of these two systems.
We find that the TIP3P water solvent and the employed force field are
able to give the correct hydration shell for residues in the interior
of the protein.
This is immediately visible from a comparison between the positions of
the crystallisation water present in the crystal structure and the
positions of water molecules with long residence time around selected
amino acids (as an example, see Figure~3~(a, b)).
We stress that the inputs for the dynamical simulations were prepared
stripping all crystallisation water molecules from the original PDB
files.
Hence, this also demonstrates that 0.5~ns is a sufficient period of
time to allow water molecules to diffuse into the inner cavities of
the protein and give the correct hydration pattern.
In the absence of the ATP ligand, water molecules are found to form a
nearly planar hydrogen-bonded network inside the binding pocket
(Figure~3~(c)).
This is consistent with the strong hydrophilicity of the back region of the 
binding pocket in an otherwise hydrophobic protein environment.
The adenine ring of ATP binds into the binding pocket of CDK2 via two
hydrogen bonds between the N1 and N6 atoms and the residues L83 and
E81, respectively, as expected from previous experimental and
theoretical investigations~\cite{Bartova_04,Otyepka_02,Schulze_96}
(Figure~3~(a, b)).
These hydrogen bonds are found to be stable and present for the whole
0.5~ns simulation.
We then perform dynamical simulations of about 1~ns for inhibitors {\bf 1}
and {\bf 2}, and of about 4~ns for inhibitors {\bf 3}, {\bf 4} and
{\bf 5}.
As mentioned above, there is no available crystal structure of
inhibitors  {\bf 2} and {\bf 5}  bound to the active CDK2/cyclin A
complex.
Therefore, in these two cases we constructed the initial inputs
by placing the inhibitors in the binding cleft of the CDK2/cyclin A
model extracted from its x-ray structure in complex with 
inhibitor {\bf 1}. 
In agreement with the crystal structures, all inhibitors are found
to form hydrogen bonds in the hydrophilic hinge region at the back
of the binding pocket.
These bonds are present for the entirety of the simulations, and the average
bond distances are in good agreement with the optimised bond lengths
obtained after full geometry minimisation within the DFT approach
(Figure~4).
Inhibitors {\bf 2}, {\bf 3} and {\bf 4} remain bound to the O atom
of E81, and the N and O atoms of L83.
In the case of inhibitor {\bf 1}, however, after $\sim$35~ps of
dynamics at 300~K the initial hydrogen bond with the O atom of L83
breaks and is immediately replaced by a new hydrogen bond with the
nearby lying O atom of H84 (Figure~5).
This binding mode is mantained for the rest of the
simulation, which is stopped after 1~ns.
Inhibitor {\bf 5} is also observed to form two strong hydrogen bonds
with the O atom of E81 and the N atom of L83, but does not form
a strong hydrogen bond with the O atom of L83.
Indeed, the donor-H-acceptor angle of this bond is $\sim$101$^\circ$,
which indicates the presence of a weaker interaction.
In addition, the O atom of L83 interacts via hydrogen bonding with
the C1 atom of the inhibitor, while an internal hydrogen-bond between the N1
atom and the O1 of the inhibitor is present in the minimised structure of
the protein/inhibitor complex (Figure~5).

Interestingly, hydrogen bonds formed with residues other than E81 and
L83 are found to be considerably less stable during the dynamics.
More specifically, the donor-acceptor distances show an intermittent
behaviour as the dynamics evolves, indicating that these bonds can be
reversibly formed and broken at room temperature.
In particular, the hydrogen bonds donated by inhibitor {\bf 3} to the
carboxylic group of the D86 residue, which are visible in the
crystallographic structure of the CDK2/inhibitor complex, are in fact
found to be present for only $\sim$21\%\ of the simulation time
(Figure~6 and Table~4).
In contrast, the corresponding interaction between D86 and inhibitor
{\bf 4} is present for the whole simulation time.
However, we observe that D86 binds to the inhibitor alternatively
switching between the OD1 and the OD2 atoms of its carboxylic
group (Figure~6).
Furthermore, we find that additional hydrogen bonds, which were not
indentified in the available x-ray structures, can temporarily form
during the dynamics.
One dramatic example is the bond which we see forming between
inhibitor {\bf 5} and the NH$_3^+$ group of K89.
In the crystal structure~\cite{Moshinsky_03}, the distance between the NZ atom
of
K89 and the N2 atom of the inhibitor is 7.1~\AA.
After about 0.3~ns of simulated dynamics at room temperature, however,
this distance suddenly decreases to about 3~\AA, indicating the formation
of a hydrogen bond (Figure~7).
This bond is then repeatedly broken and reformed as the dynamics proceeds,
and is present for $\sim$20\%\ of the time over a total simulated time of
more than 4~ns (see next section).
In an analogous fashion, K89 is observed to form intermittent hydrogen
bonds also with inhibitors {\bf 4} and {\bf 5} (Figure~6).
Intermittent hydrogen bonding appears thus to be a key element of the
interaction between CDK2 and its inhibitors, and need to be taken into
account when calculating the binding energies of the protein/inhibitor
complexes.

\subsection{Calculation of binding energies from the AMBER structures}

In order to account quantitatively for the intermittent hydrogen bonds, 
we first perform a full structural minimization within the classical force field
starting from a snapshot of the dynamics when all hydrogen bonds are present at 
the same time.
A model is then extracted from the minimised structure,
including only the inhibitor and those amino acids which are in direct
hydrogen-bond contact with the inhibitor.
As done in the case of the models extracted from the crystal
structures, the side chains of the selected amino acids are replaced
by methyl groups whenever they do not participate in any hydrogen
bond.
The geometry of this small inhibitor/pocket model system (HB model) 
is fully minimised using the PW DFT approach, keeping the
C$_{\alpha}$ atoms of the backbone fixed.
Finally, the binding energy of the inhibitor to the small system
is calculated as in Section 3.2 ($\Delta E_g$(HB) values in Table~3).
This minimal pocket model is then embedded in a larger model
comprising all amino acids and water molecules within 7.0~\AA\ 
from the inhibitor, in the geometry obtained after full minimisation
within the force field potential.
An additional force field minimisation of this larger pocket model
is performed keeping fixed the $C_{\alpha}$ atoms and all atoms
of the smaller HB model.
This is necessary to relax any possible steric clash between the added
amino acids and water molecules and the elements of the HB model
after the minimisation at the DFT level~\cite{Phe_82}.
After this minimisation, the binding energy within the larger pocket model is
calculated as the difference in the DFT total energy between this
system, the isolated inhibitor, and the isolated pocket model, without
relaxing the geometries of the individual components
($\Delta E_g$(7\,\AA) values in Table~3).

We note that all values of $\Delta E_g$(7\,\AA) are smaller
than the corresponding values of $\Delta E_g$(HB).
This indicates that the direct hydrogen bond interactions between the
inhibitors and the protein are substantially screened by the presence
of the surrounding amino acids and solvation water molecules.
Interestingly, when all water molecules are stripped off from the 7\,\AA\
model, the computed  binding energies  ($\Delta E_g$(7\,\AA\ dry) in
Table~3) are slightly larger than the $\Delta E_g$(HB) values,  as found
in section 4.2 for the systems extracted from the crystal structures.
Furthermore, in the case of inhibitor {\bf 5}, we computed the binding
energies of systems composed of all amino acids of the HB model and all
the water molecules within 4.0\,\AA,  6.0\,\AA, and 7.0\,\AA\ from the
inhibitor. 
The computed $\Delta E_{g}$values (in kcal/mol) for the HB model and
these three systems are -50.3, -50.0, -49.6, -49.1, indicating that
the presence of water molecules alone does not contribute substantially
to the observed screening of the interactions passing from the HB to
the 7\,\AA\ systems.
The combination of these two results leads to the hypothesis that the observed
screening effect must arise from the {\em  simultaneous} presence of both
the surrounding amino acids and the solvation water molecules.

Indeed, the binding energy of inhibitor {\bf 5} calculated adding to the
HB system residues 10 and 86, as well as the backbone of residues 84 and
85, is -44.1~kcal/mol, while the binding energy value of the same system after
adding a single water molecule which bridges the NH$^{3+}$ group of K89
and the peptide group between residues 84 and 85 (Figure 8) is 
decreased dramatically to -39.1~kcal/mol.
This drop in binding energy is associated with a substantial change 
in  the polarisation of the hydrogen bonds upon inclusion of the further
protein residues and the water molecule, as can be seen by
analysis of the atomic charges which best fit the electrostatic 
potential computed from first principles.
Namely, calculated atomic charges on the acceptor N atom of {\bf 5}
changes from  -0.23 in the HB model to 0.01 in the presence of the
additional residues,  to 0.10 after adding the water molecule, while
the charge on the donor H atom of K89 changes from 0.19 to -0.02 to
-0.26.
This is indicative of profound changes in the local electrostatics, and 
thus in the hydrogen bond strength, due to rearrangements of electronic
charge in the three systems considered.
This result highlights the importance of calculating binding energies at a
quantum level of accuracy, and reveals water-mediated polarization effects~%
\cite{Polarization} as a novel feature of the interaction between CDK2 
and its inhibitors.

The values of the binding energy reported in Table~3 represent
an upper limit to the actual binding energy of the system.
In fact, since many of the hydrogen bonds are repeatedly broken and
formed during the motion of the protein at room temperature, their
contributions to the binding energy need to be ``weighted'' by the
fraction of time for which they are present (Table~4).
Taking as an example inhibitor {\bf 3}, a total of six hydrogen bonds
are formed: the triplet of stable bonds with E81 and L83, a pair of
intermittent bonds with D86, and an intermittent bond with K89.
Let $t^{tri}$  $t^{tri,86}$, $t^{tri,89}$, and $t^{tri,86,89}$ be the
fractions of time for which the bonds with the triplet only, the
triplet and residue 86, the triplet and residue 89, and the triplet,
residue 86 and residue 89 are present, respectively.
These values can be extracted from an analysis of the variation of
the hydrogen-bond distances and angles during each simulation.
For the presence of a bond we define a cut-off distance of
4.0~\AA\ between donor and acceptor, and a minimum angle
of 90$^\circ$ between donor, hydrogen, and acceptor.
We stress the importance of including a condition on the hydrogen bond
angle, since the hydrogen bond strength is considerably reduced
when this angle deviates from the ideal linear conformation~%
\cite{Ireta_04}.
In the case of inhibitor {\bf 3} we obtain $t^{tri,86} = 0.13$,
$t^{tri,89} = 0.18$, and $t^{tri,86,89} = 0.08$, and
$t^{tri} = 0.61$.

Now we need to split the computed values of $\Delta E_g$ into the
individual contributions corresponding to each binding mode.
We define $\Delta E_g^{tri}$, $\Delta E_g^{tri,86}$, $\Delta E_g^{tri,89}$,
and $\Delta E_g^{tri,86,89}$ as the binding energies of the inhibitor to
the pocket in the presence of the hydrogen bond triplet only, the
triplet and the residue 86, the triplet and residue 89, and the
triplet together with both residue 86 and 89, respectively.
Each of these contributions are calculated separately via static total
energy calculations of small models extracted from the geometry of the
HB system previously obtained.
At this point, we make the assumption that the ratios between the
contributions of the individual bonding modes are the same in the
HB and in the 7\AA\ pocket models.
In other words, the individual contributions calculated taking into
account only the minimal pocket model are scaled by the same factor
$\Delta E_g({\rm 7\AA})/\Delta E_g({\rm HB})$ (0.85 in 
the case of inhibitor {\bf 3}).
Within this assumption, we can finally calculate the binding energy
$\Delta E_g$ as follows:
\begin{equation}
 \Delta E_g =  \frac{\Delta E_g(\mathrm{7\AA})}{\Delta E_g(\mathrm{HB})}\cdot
               \left( t^{tri}\Delta E_g^{tri} + t^{tri,86}\Delta E_g^{tri,86} +
               t^{tri,89}\Delta E_g^{tri,89} + t^{tri,86,89}\Delta E_g^{tri,86,89}\right).
\end{equation}
The computed values of binding energy of the individual binding modes
of all five inhibitors are reported in Table~5.

Since explicit water molecules are included in our models, the values
of $\Delta E_g$ for the five inhibitors calculated with the procedure
described above already contain the enthalpic contributions to
the solvation of the $PL$ and the $P$ systems.
The remaining contributions to the solvation free energy, i.e. the surface
area contributions of the $PL$ and $P$ models, as well as the solvation
free energy of the isolated ligands, are calculated within the GBSA method
to obtain the solvation free energy differences $\Delta G_{solv}^{wat}$,
where the superscript indicates that these values refer to the models including
explicit water solvent (reported in Table~3).
The obtained enthalpic and solvation contributions can now be used to 
compute a final set of free energy differences $\Delta\Delta G$ (Table 6).
The inferred $K_{i}$ ratios are now found to agree with the experimentally 
measured values within one order of magnitude for all the inhibitors 
considered, including the delicate case of inhibitor {\bf 5}.

\section{Conclusions}

\subsubsection*{Binding modes and binding energies}

The main conclusion that can be drawn from our investigations
is that the potency of inhibition of small molecules which
bind to the ATP pocket of CDK2 is largely dependent on the
dynamical nature of the protein/inhibitor interactions.
In particular, the motion of both the inhibitor and
of particular residues within the binding pocket is responsible for
the intermittent formation and rupture of hydrogen bonds.
Such effects are not easy to identify by using only static methods for
structure investigation, such as x-ray crystallography.
Moreover, even if the orientation of the inhibitor in the binding
pocket remains the same, differences in the conformation of the backbone
of the protein can lead to substantial differences in the magnitude of
the calculated hydrogen bond strengths (cf. Tables~1 and~3).
The dynamical nature of the hydrogen bond interactions is likely to be
the main reason for the quantitative disagreement between the
experimental binding energy values and the values obtained using
scoring functions after rigid docking simulations (see Table~2
and Ref.~\cite{Otyepka_00}).
Notably, the binding free energy differences calculated with this
method are underestimated with respect to the experimental values,
which is consistent with the neglection of additional intermittent interactions
within the protein active site.
On the other hand, the scoring functions employed are carefully
parametrised to account for hydrophobic and solvation effects in a
simple but robust manner.
Indeed, they are capable of giving a better qualitative estimate of
the rank of potencies between the considered inhibitors than
simplistic DFT binding energy calculations after geometry optimisation
of protein/inhibitor complexes extracted from the crystal structures,
in the absence of explicit solvent molecules (Table~1).

Our results stress the importance of including a large number of amino
acids in the models for the ATP binding pocket (see Figure~2), since
long range electrostatic effects contribute substantially to the
computed binding energies.
In general, when no explicit water molecules are included in the
calculations, the binding energy increases with increasing number
of amino acids included in the calculations (Figure~2 and Tables~1
and~3), converging within 0.15~kcal/mol when all amino acids within
7~\AA\ from the inhibitor are considered.
A major change to the computed binding energy is due to the explicit
presence of water molecules within the binding pocket. 
In this case hydrogen bond interactions are substantially screened due to
water-mediated hydrogen bond depolarisation effects, so that the
binding energy is reduced up to $\sim$35~\% of its original value
(Tables~3 and~5).
Since relatively few water molecules can be identified
by crystallographic methods, this stresses the importance
of using MD methods to obtain a correct solvated model of
the protein/ligand complex.
Moreover, given the complex nature of the screening effects
observed, this indicates that calculations of binding energies 
necessarily require the use of quantum techniques like DFT.

The main limitation of MD methods is that the accessible simulation 
time is very short.
Therefore, the fractions of time associated with each individual
binding modes which enter into the calculations of the relative free
energy of binding should be considered as indicative
values~\cite{Boltz_fac}.
Given this limitation, the approximations made on the solvation 
and entropic terms contributing to the binding energies appear to be
less severe.
While gas-phase entropic terms (vibrational, internal rotational and
conformational) necessarily need to be explicitly considered in
order to compute {\em absolute} values of binding free energies, this
is not necessary when comparing {\em relative} values of binding free
energies among different inhibitors of the same protein~\cite{Wang_01}.
As mentioned in Section~3, this approximation is justified
by the chemical similarities of the systems considered.
Also the underestimation of direct van der Waals interactions
within the DFT techniques employed appears not to be crucial
in the present case.
Dispersion forces can represent large fractions of absolute
binding energies in biological systems~\cite{Hobza_05,Elstner_01,Barrat_05}, 
and are indeed of the order of 45~kcal/mol in our minimised CDK2/ligand 
complexes (i.e. at the maximum van der Waals contact between the 
hydrophobic and aromatic residues in the binding pocket and the ligands).
However, when the differences of van der Waals energies are computed 
as averages of relatively long dynamic runs at 300~K, they are
found to be the same for all the inhibitors considered in this 
work (within the error bar of a few kcal/mol intrinsic to the MM 
method), and thus they do not contribute substantially to the 
calculated $\Delta \Delta G$ values.

As far as the calculation of solvation free energies is concerned, 
this is in general a very difficult task, requiring in principle 
accurate formalisms and an explicit treatment of solvent 
molecules~\cite{Mancera_wat}.
Here we have used the widely used continuum GBSA model~\cite{Sitkoff_94} to 
compute the solvation free energies (apart from the final set of calculations, 
where explicit water molecules are included in the $PL$ and $P$ models, 
see Section 4.1), expecting errors in the {\em differences} between 
solvation free energies calculated within the same method to be
negligible compared with all other approximations.

\subsubsection*{Specificity of inhibition}

Considering the approximations inherent to the methods employed
and the limited statistics accessible to atomistic MD simulations, the
agreement between calculated and measured relative potencies is
remarkably good.
It has to be noted, however, that direct calculations of inhibition constants is
still out of the reach of this and analogous computational schemes.
The power of our method lies in the ability of predicting and quantifying
all interactions present between a given inhibitor and, in this case, CDK2.
In particular, in the case of inhibitor {\bf 5} analysis of the
computed hydrogen bond strengths immediately revealed that a strong
interaction was absent from the previously considered structural
model.
With the help of dynamical simulations, this interaction has been
identified in the form of an intermittent hydrogen bond with Lys89,
which was not visible in the crystal structure~\cite{Moshinsky_03}.
The same residue has been observed to form similar intermittent
interactions also with inhibitors {\bf 3} and {\bf 4}.
The presence of such interactions is particularly important,
since Lys89 is a residue specific to CDK2.
In CDK4, for instance, a much shorter threonine residue is present in the
same position (Thr89), which is not expected to be able to form hydrogen
bonds with small inhibitors, unless they are specifically modified
for this purpose~\cite{Honma_01}.
Therefore, there is an increase in binding energy between the
inhibitors and the protein which is specific to CDK2, and is
not expected to be present in the case of CDK4.
This is consistent with the observed, yet so far unexplained selectivity
patterns of inhibitors {\bf 3}~\cite{Davies_02} and {\bf5}~%
\cite{Moshinsky_03}.

Recent inhibitor profiles have targeted the Lys~89 residue for
this very reason, placing a highly hydrophilic group at a
favourable position~\cite{Hamdouchi_05} .
However, we note that the gain in binding enthalpy from adding
hydrophilic groups targeting Lys89 is counterbalanced by the
penalty of an increased solvation energy.
Inhibitor {\bf 5} is an interesting case, having a considerably
lower solvation energy than other pyridine and pyrimidine based
inhibitors, yet still being capable of interaction with Lys89.
In principle, it is also conceivable that Lys89 could form an intermittent
hydrogen bond with the N1 atom of inhibitor {\bf 1}, which could help
explain why also in the case of this inhibitor a moderate selectivity
against CDK4 has been observed~\cite{Davies_02}.
On the other hand, this selectivity could also arise from factors not
directly connected to local interactions within the ATP binding
cleft, or from more subtle effects, e.g. differences between the
hydrophobicity patterns of the two proteins~\cite{Mancera_06}.

\subsubsection*{Summary}

The relative differences of binding free energies $\Delta\Delta G$
between five inhibitors of CDK2 have been calculated by using a
combination of linear-scaling DFT techniques and classical MD
simulations.
The $\Delta\Delta G$ values obtained after direct structural
optimisation of protein/inhibitor complexes extracted from the
available crystal structures are found not to correlate well 
with the measured potencies of inhibition.
As a comparison, extensive docking simulations followed by free
energy calculations using scoring function methods provided a better
qualitative estimation of the rank of potencies among the inhibitors,
although a quantitative agreement could not be achieved.
Molecular dynamics simulations using classical force-field techniques
revealed that some of the hydrogen bonds formed between the inhibitors
and the ATP binding cleft of CDK2 are in fact intermittent, being
repeatedly broken and formed at room temperature.
These dynamical effects are taken into account in the O(N) DFT binding
energy calculations by weighting each interaction by the percentage
of the time for which it is actually present.
Moreover, a combined effect due to the presence of solvation water
molecules and neighbouring amino acids leads to a substantial
screening of the protein/ligand hydrogen bond interactions, which 
can be accurately quantified using DFT techniques.
Despite the approximate nature of the methods and the limited
statistics available, $\Delta\Delta G$ values calculated taking these
effects into account show a much better agreement with the
experimentally measured potencies of these inhibitors.
Finally, the simulations reveal the presence of intermittent
hydrogen-bond interactions between the inhibitors and the
Lys89 residue, which have not been identified in the
crystallographic studies performed to date.
As Lys89 is a CDK2 specific residue, while in CDK4 a much shorter Thr
residue is present in position 89, the existence of such an interaction 
offers an explanation for the observed, yet so far not understood specificity 
of some of the inhibitors for CDK2 against CDK4.

Taken together, our findings reveal new factors underlying the relationship
between the structure and activity of CDK inhibitors.
They demonstrate that polarization effects arising from the specific hydration
patterns in the ATP-binding pocket of CDK2 contribute to the differential
potency of its ihibitors.
Such polarization effects may be a general feature in other kinase-inhibitor
complexes, which must be taken into account for rational drug design.
Our results also demonstrate a contribution of intermittent hydrogen bonding
interactions, not previously recognized in static crystallographic studies, to the
potency of CDK2 inhibitors and their relative selectivity over CDK4.
Finally, this work provides a generally applicable computational method which
can be adapted to the investigation of other kinase-inhibitor structures, and which
could be a useful tool in rational drug design.

\section{Computational Details}

\subsubsection*{Density Functional Theory calculations}

Quantum mechanical calculations of total energies were performed both
with traditional plane-wave (PW)~\cite{Payne_92}, and with novel
linear-scaling (O(N))~\cite{Soler_02,Skylaris_05}
implementations of Density Functional Theory (DFT),
using the PBE gradient corrected exchange-correlation
functional~\cite{Perdew_96}.
The \textsc{Castep}~\cite{Segall_02} code was employed for the PW
calculations of smaller systems, using ultrasoft pseudopotentials~%
\cite{Vanderbilt_90}, and expanding the wave functions at the $\Gamma$-point
of the Brillouin zone up to a kinetic energy cut-off of $450$~eV.
The modelled molecules were placed in cubic simulation cells with
edge length of 25{\AA}, which was sufficient to avoid any spurious
interactions between the system and its periodicallly repeated images
in all cases considered.
The computing cost of the PW approach increases with the cube
of the number of atoms and becomes prohibitively expensive for systems
with more than a few hundreds of atoms.
In contrast, O(N) methods have recently been developed where the
increase in computing cost is linear with the number of atoms, and
allow us to extend the DFT calculations to much larger systems.
O(N) DFT calculations were performed using both the widely used
\textsc{Siesta} code~\cite{Soler_02} and the newly developed
\textsc{Onetep} code~\cite{Skylaris_05}.
In both cases the interactions between electrons and nuclei were
described by norm-conserving pseudopotentials~\cite{Hamann_79,Troullier_91}.
\textsc{Siesta} calculations were performed by placing the systems
in cubic supercells with edge lengths of up to 65~\AA, avoiding any
spurious interactions between the simulated system and its periodically
repeated images.
The employed basis set consists of a DZP basis of localized atomic
orbitals which were variationally optimized to accurately describe
hydrogen bonds in biomolecular systems\cite{Anglada_02}.
Due to the atomic basis set, calculations of binding energies
needed to be corrected from the so-called basis set superposition
error (BSSE) using the counterpoise method of Boys and Bernardi~%
\cite{Boys_70}.
In the \textsc{Onetep} code the DFT equations are solved using
a  minimal set of strictly localised functions which are 
optimised {\em in situ} in a basis set which is formally 
equivalent to a plane wave expansion~\cite{Skylaris_02-1}.
Therefore, the same levels of accuracy and rapidity of convergence
of traditional PW codes are achieved~\cite{Mostofi_03}, and there
is no superposition error to be corrected during the calculation
of binding energies~\cite{Haynes_05}.
The \textsc{Onetep} calculations were performed using a kinetic energy
cut-off of 600~eV and placing the systems in a cubic simulation
supercell with edge length of about 80~\AA.
In all calculations charge neutrality of the systems was imposed by
protonation or deprotonation of residue groups of the protein models
which were not involved in any direct interaction with the
inhibitors.

We note that our calculations provide an important comparison between a
well established, but computationally costly, DFT implementation, and
two recent linear-scaling methods for the quantum mechanical treatment
of very large systems (up to thousands of atoms) based on two very
different mathematical formalisms.
Our results show that the differences in the computed binding energies
between the three different codes are less than 1.5~kcal/mol for
systems including about 1000 atoms (see Section 4.1), i.e. within the
error bar associated with the other approximations employed (supercell
method, pseudopotentials, exchange-correlation functionals).
For instance, the binding energy values of one of the inhibitors
calculated with the \textsc{Castep}, \textsc{Siesta}, and
\textsc{Onetep} codes are -16.2~kcal/mol, -16.1~kcal/mol, and
-16.4~kcal/mol, respectively (another comparisons between the three
codes  have been presented in Section~3.1).
Hence the results obtained with any of the codes are entirely
equivalent at the standard accuracy level reached by DFT techniques.
Nevertheless, to ensure full consistency of the presented results, 
the reported values of the differences in binding energies 
are the ones calculated with the \textsc{Onetep} code.
The large volume calculations necessary at the exploratory and set-up
stages of the work (concerning size convergence, tests on charged
states, choice of protonation sites, etc.) were performed using the
\textsc{Siesta} code.

\subsubsection*{Classical Force-field Simulations\label{forcefield}}

The \textsc{Amber}~7.0~\cite{amber7} suite of programmes was used for
all the classical force-field simulations.
Point charges on the atoms of the inhibitors were calculated as a
best-fit to the electrostatic potentials computed within the DFT
formalism in a 1.0~\AA\ thick spherical region outside the Van der
Waals radii of the atoms.
All other parameters for the inhibitors were determined using the
Antechamber facility making use of the gaff force-field provided.
Solvation free energies were calculated within the standard pairwise
Generalised Born model, taking into account entropic contributions
using the the LCPO method to calculate the molecular surface 
area.
The classical simulations with explicit solvent were carried out using
the ff99~\cite{Wang_00} force field based on the parameters of Cornell
et~al.~\cite{Cornell_95}, and the TIP3P water solvent.
The models for the classical simulations were prepared by reading into
the Leap module of \textsc{Amber} the crystal structures of the
protein/inhibitor complexes after having stripped all crystallographic
water molecules and counterions present.
Each model was then protonated, solvated in a tetragonal box with
edge lengths ensuring a mininum 20~\AA\ distance from the system and
its periodically repeated images, and neutralised by adding chloride
counterions.
The dynamical simulations were performed following a standard
three-step equilibration procedure:
(i) The system was first fully minimised to release any unwanted steric
clashes, and then the solvent was equilibrated in a 0.015~ns run slowly
increasing the temperature to 300~K at the constant pressure of 1~atm,
keeping the solute frozen.
(ii) A series of 5 minimisation runs was then performed, in which the
harmonic constraints on the solute were progressively released and
eventually eliminated.
(iii) Subsequently, the system was equilibrated in a 0.05~ns run to
reach conditions of constant temperature (300~K) and pressure (1~atm).
After equilibration, we carried out the production run for a time
which varied from 0.5 to 4.0~ns, depending on the system considered.
Only the production runs were taken into account in the analyses of the
hydrogen bond patterns (Section~5.1).
We used an Ewald cut-off sphere of 12.0~\AA\ and  a timestep of 2~fs for
the integration of the equations of motion, employing the SHAKE algorithm
to keep fixed all bonds involving H atoms.
A harmonic constraint of 5.0~kcal/mol/\AA$^{2}$ was kept on the amino acids
immediately before and after any protein sequences unavailable in the
crystal structure (typically in the region from residue 36 to residue
43 of the CDK2/inhibitor complexes).

\subsubsection*{Van der Waals dispersion interactions}

The standard DFT technique employed systematically underestimates
direct interactions due to dispersion forces.
We thus carefully checked that the dispersion interactions between the 
protein and the ligands in the binding pocket are roughly the same for 
all inhibitors considered, so that they do not contribute significantly 
to the computed free energy differences $\Delta\Delta G$.
An estimate of the maximum dispersion interactions was obtained from the
values of the Van der Waals energies of the minimised $PL$ complexes and 
the isolated $P$ and $L$ systems in the same atomic configurations as 
in the $PL$ complex, computed with the classical Amber force field in an
implicit solvent.
For the NU2058, NU6027, SU9516, 9d-NU6027 and the NU6102 inibitors, 
we compute net Van der Waals interaction energies of 41.4, 42.0, 40.1, 54.2
and 55.8 kcal/mol, respectively. 
These large values are consistent with other quantum chemical and
empirical calculations of dispersion energies in aromatic systems~%
~\cite{Hobza_05,Elstner_01}.
We note, however, that the dispersion interactions are considerably
affected by the brownian motion of the ligand and of the residues within 
the binding pocket in water solution at 300~K.
We therefore computed the average of the Van der Waals binding energies
over the configurations of the $PL$ complexes during the dynamics at 300~K 
in an explicit solvent for the most weakly and most strongly bound 
inhibitors, i.e. SU9516 and NU6102. 
The differences $\Delta E_{vdw}(PL) - \Delta E_{vdw}(P) - \Delta E_{vdw}(L)$
between those inhibitors were found to be less that $\sim$4~kcal/mol, 
which is within the error bar of the method.
We therefore conclude that direct dispersion interactions do not contribute
in a substantial way to the different potencies among the inhibitors, so
the differential values of binding energy can be safely computed within
the DFT scheme.

\subsubsection*{Scoring Functions calculations}

Docking simulations were carried out using a recently introduced
quantum stochastic tunnelling docking method~\cite{Todorov_03,Mancera_04}.
This is a very efficient hybrid optimisation method that allows for
the quantum and stochastic tunnelling through high energy barriers and
the non-local exploration of the potential energy surface.
Two hundred simulations were performed for each ligand-protein complex
under the same conditions as described in~\cite{Mancera_04}, using the
PLP scoring function~\cite{Gehlhaar_95} to represent the potential energy
surface.
A large number of binding modes were generated, which were then
re-scored using the B\"ohm~\cite{Boehm_94} and ScreenScore~\cite{Stahl_01}
scoring functions.
The free energy of binding (for each scoring function) of each ligand-protein
complex reported corresponds to the best energy found amongst all generated
binding modes within 1.0~\AA\ RMSD of the binding mode visible in the
crystal structure.
This precaution ensured that the best estimate of the free energy binding
corresponded to the experimentally observed binding mode.

\section{Acknowledgements}
The authors are indebted with J.A. Endicott and M.E.M. Noble for many
valuable discussions.
M. Segall, C. Molteni, C. Pickard are acknowledged for useful
suggestions.
C.-K.~S. would like to thank the Royal Society for a University
Research Fellowship.
L.C.C. acknowledges support by the Alexander von Humboldt foundation.
Computing resources were provided by the Cambridge-Cranfield High
Performance Computing Facility and the HPCx computing facilities
through the UKCP Consortium.
This work is supported by EPSRC (Grant GR/S61263/01).

\cleardoublepage
\newpage

\section*{Tables}
\vspace*{2cm}

\begin{table}[h]
\centering
\begin{tabular}{| c | l | c | c | c | c | c | c |}
\hline
Inhibitor &  $K_i$~($\mu$M) exp. &   $\Delta E_g$(HB)  &  $\Delta E_g$(7\,\AA) &
              $\Delta G_{solv}$  &  $\Delta\Delta G$ & $\Delta\Delta G$
exp.\\
\hline
{\bf 1}  & {\em (1.2$\pm$0.3)E+1} &  -20.0  &  -18.3  &  19.4  &  0.0 &
{\em  0.0} \\
{\bf 2}  & {\em (1.3$\pm$0.2) }   &  -18.5  &  -18.4  &   21.4  & +1.6 &
{\em -1.4} \\
{\bf 5}  & {\em (3.1$\pm$0.6)E-2} &  -14.6  &  -15.1  &  19.4  & +2.9 &
{\em -3.7} \\
{\bf 4}  & {\em (2.4$\pm$0.8)E-2} &  -37.1  &  -43.4  &  41.4  & -3.4 &
{\em -3.8} \\
{\bf 3}  & {\em (6.0$\pm$0.5)E-3} &  -43.3  &  -48.2  &  32.7  & -14.1 &
{\em -4.7} \\
\hline
\end{tabular}
\caption{Binding energies, solvation energies, and relative binding
free energy differences (all in kcal/mol) of five CDK2 inhibitors
calculated at the DFT level of theory compared to the experimentally
measured inhibition constants and relative difference of free energy
of binding (in italics). The binding energy values are obtained
starting from the available crystal structures after geometry
optimisation of the hydrogen bond distances of the protein/ligand
complexes.}
\end{table}

\begin{table}[h]
\centering
\begin{tabular}{| c | l | c | c | c | c | c | c |}
\hline
 Inhibitor &  PLP  &  B\"ohm  & ScreenScore  & $\Delta\Delta G$ exp.  \\
\hline
{\bf 1}  &    0.0   &   0.0  &    0.0    &  {\em  0.0}   \\
{\bf 2}  &   +2.1   &  +0.5  &   -0.3    &  {\em -1.4}   \\
{\bf 5}  &   -7.2   &  -1.5  &   -1.0    &  {\em -3.7}   \\
{\bf 4}  &  -35.5   &  -1.4  &   -1.4    &  {\em -3.8}   \\
{\bf 3}  &  -37.0   &  -3.1  &   -3.7    &  {\em -4.7}   \\
\hline
\end{tabular}
\caption{Relative differencies in the free energies of binding
calculated with three different sets of scoring functions after
extended docking simulations, compared with the experimental values
(in italics).  PLP values are in arbitrary energy units, B\"ohm and
ScreenScore values are in kcal/mol.}
\end{table}

\begin{table}
\centering
\begin{tabular}{| c | c | c | c | c |}
\hline
Inhibitor & $\Delta E_g$(HB) &  $\Delta E_g$(7\,\AA)  &  $\Delta E_g$(7\,\AA\ dry) 
          & $\Delta G_{solv}^{wat}$    \\
\hline
{\bf 1}  &  -16.2  &  -12.5  &  -17.4  &  10.7 \\
{\bf 2}  &  -17.3  &  -13.8  &  -18.5  &  11.4 \\
{\bf 5}  &  -50.3  &  -34.3  &  -51.3  &   9.7 \\
{\bf 4}  &  -57.5  &  -36.7  &  -59.1  &  15.4 \\
{\bf 3}  &  -74.6  &  -63.6  &  -81.7  &  19.4 \\
\hline
\end{tabular}
\caption{Binding energies (kcal/mol) calculated at the DFT level of theory after 
 MD simulations and geometry minimisation of the protein/ligand complexes
using the \textsc{Amber} force-field, and DFT optimisation of the hydrogen bond
distances, considering three models of the binding pocket as described
in the text, Section 4.1., as well as the contribution to the binding free
energy due to solvation, $\Delta G_{solv}^{wat}$. }
\end{table}

\begin{table}
\renewcommand{\baselinestretch}{1.3}\small\normalsize
\label{tab:hbonds}
\centering
\begin{tabular}{| c | c | c | c | c | c |}
\hline
Inhibitor &  H-bond &    H-bond    & length  (\AA)    & length (\AA)     &   t    \\
             &  (x-ray) &    (MD)    & (x-ray)  & (DFT)  & 
(\%)
 \\
\hline
{\bf 1}   &  E81(O)  &    E81(O)     &    2.73     &    2.86    &   100  \\
          &  L83(N)  &    L83(N)     &    3.00     &    3.61    &   100  \\
          &  L83(O)  & {\em H84(O)}  &    2.96     &    2.95    &   100  \\
\hline
{\bf 2}   &  E81(O)  &    E81(O)     &    2.79     &    2.81    &   100  \\
          &  L83(N)  &    L83(N)     &    2.95     &    3.12    &   100  \\
          &  L83(O)  &    L83(O)     &    2.83     &    2.85    &   100  \\
\hline
{\bf 5}   &  E81(O)  &    E81(O)     &    3.00     &    2.77    &   100  \\
          &  L83(N)  &    L83(N)     &    2.94     &    2.92    &   100  \\
          &  L83(O)  &    L83(O)     &    4.75     &    3.06    &   100  \\
          &    -     & {\em K89(NZ)} &             &    2.77    &    20  \\
\hline
{\bf 4}   &  E81(O)  &    E81(O)     &    3.03     &    2.79    &   100  \\
          &  L83(N)  &    L83(N)     &    3.24     &    3.54    &   100  \\
          &  L83(O)  &    L83(O)     &    2.92     &    3.07    &   100  \\
          &  D86(OD2)& D86(OD1,OD2)  &    2.79     &    2.70    &   100  \\
          &    -     & {\em K89(NZ)} &             &    2.53    &    27  \\
\hline
{\bf 3}   &  E81(O)  &    E81(O)     &    2.97     &    2.76    &   100  \\
          &  L83(N)  &    L83(N)     &    3.15     &    3.27    &   100  \\
          &  L83(O)  &    L83(O)     &    2.97     &    2.99    &   100  \\
          &  D86(N)  &    D86(N)     &    3.74     &    3.18    &    21  \\
          &  D86(OD2)& D86(OD1,OD2)  &    2.72     &    2.58    &    21  \\
          &    -     & {\em K89(NZ)} &             &    2.65    &    26  \\
\hline
\end{tabular}
\caption{Hydrogen bonds visible in the crystal structures (x-ray) and
during force field MD simulations (MD). Donor-acceptor pairs and hydrogen 
bond distances after DFT optimisation are reported together with the time 
percentage for which each bond is present in the MD simulations. 
Residues making hydrogen bonds not visible in the x-ray structure are 
reported in italics. OD1 and OD2 are the oxygen atoms of the carboxyl 
group of aspartic acid; NZ is the nitrogen atom of the NH$_3^+$ group of 
lysine; O and N are the peptide oxygen and nitrogen atoms of the 
indicated residues.}
\end{table}

\begin{table}
\centering
\begin{tabular}{| c | c | c | c | c | c |}
\hline
Inhibitor & $\Delta E_g$(7~\AA)/$\Delta E_g$(HB)  &  $\Delta E_g^{tri}$   &
            $\Delta E_g^{tri,86}$  &   $\Delta E_g^{tri,89}$    & 
$\Delta E_g^{tri,86,89}$
\\
\hline
{\bf 1}  &   0.772  &  -16.2  &   -     &     -    &    -    \\
{\bf 2}  &   0.798  &  -17.3  &   -     &     -    &    -    \\
{\bf 5}  &   0.682  &  -17.6  &   -     &  -50.3   &    -    \\
{\bf 4}  &   0.638  &  -12.8  &  -23.5  &     -    &  -57.5  \\
{\bf 3}  &   0.853  &  -18.3  &  -42.0  &  -47.0   &  -74.6  \\
\hline
\end{tabular}
\caption{Ratios between the binding energies calculated in the HB 
and 7\,\AA\ models and binding energy values (kcal/mol) of systems
composed by the inhibitors and different subsets of amino acids, 
as described in the text, Section 4.1.}
\end{table}

\begin{table}
\centering
\begin{tabular}{| c | c | c |c| c|}
\hline
Inhibitor & $\Delta\Delta G$  &  $\Delta\Delta G$ exp. & $ e^{\Delta\Delta G / k_B T}$
          & $K_i$/$K_i^{NU2058}$ exp.  \\
\hline
{\bf 1}   &  0.0  &  {\em 0.0}  & 1.0       & {\em 1.0 }                \\
{\bf 2}   & -0.6  & {\em -1.4}  & 3.8\,E-1  & {\em (1.1 $\pm$ 3.2)E-1}  \\
{\bf 5}   & -4.9  & {\em -3.7}  & 3.5\,E-4  & {\em (2.6 $\pm$ 0.8)E-3}  \\
{\bf 4}   & -3.6  & {\em -3.8}  & 2.9\,E-3  & {\em (2.0 $\pm$ 0.8)E-3}  \\
{\bf 3}   & -5.3  & {\em -4.7}  & 1.8\,E-4  & {\em (5.0 $\pm$ 1.3)E-4}  \\
\hline
\end{tabular}
\caption{Computed differences of binding free energy (kcal/mol) and relative
potencies compared with experimental values (in italics) obtained from
the measured inhibition constants.}
\end{table}

\cleardoublepage
\newpage

\section*{Figure captions}

\begin{figure}[h]
\caption{The five CDK2 inhibitors considered in this study, NU2058
({\bf 1}), NU6027 ({\bf 2}), NU6102 ({\bf 3}), 9d-NU6027 ({\bf 4}),
SU9516 ({\bf 5}); the natural CDK2 ligand, ATP ({\bf 6}); and
staurosporine ({\bf 7}).}
\end{figure}

\begin{figure}[h]
\caption{Convergence with CDK fragment size of binding energy values
(kcal/mol) calculated using two different O(N) DFT codes for the case
of staurosporine bound to the ATP pocket of CDK2.}
\end{figure}

\begin{figure}[h]
\caption{Comparison between the position of selected solvation water
molecules visible in the crystal structure of the CDK2/ATP complex (a)
and the position of water molecules with long residence time in an MD
simulation of the same system (b). Corresponding molecules are
highlighted in green, and the hydrogen bonds made by ATP in the
binding cleft are shown with dotted lines. In (c) and (d) a comparison
is shown between the planar network of water molecules and the
position of the ATP ligand within the binding cleft of CDK2.}
\end{figure}

\begin{figure}[h]
\caption{Evolution of the hydrogen bond distance between inhibitor {\bf
3} and the peptide nitrogen of Leu 83 during an MD dynamical simulation,
compared with the distance calculated at the DFT level (horizontal line).}
\end{figure}

\begin{figure}[h]
\caption{Binding modes of inhibitors {\bf 1} (a) and {\bf 5} (b) in
the ATP binding cleft of CDK2 as obtained from MD simulations. The
formed hydrogen bonds are indicated with dotted lines.}
\end{figure}

\begin{figure}[h]
\caption{Evolution of the distances of hydrogen bonds formed between
inhibitors {\bf 4} (left) and {\bf 3} (right) and residues D86 (top) and
K89 (bottom) of CDK2.}
\end{figure}

\begin{figure}[h]
\caption{Evolution of the distance of the hydrogen bond formed between
inhibitor {\bf 5} and residue K89 of CDK2.}
\end{figure}

\begin{figure}[h]
\caption{A water molecule (indicated with an arrow) bridging the
NH3$^{+}$ group of K89 with the peptide O atom of H84 near the SU9516
inhibitor. The calculated binding energy decreases by 5~kcal/mol when
this single water molecule is included in the DFT total energy
calculations (see text, Section 4.1).}
\end{figure}

\cleardoublepage

\begin{center}
\includegraphics[width=8.0cm]{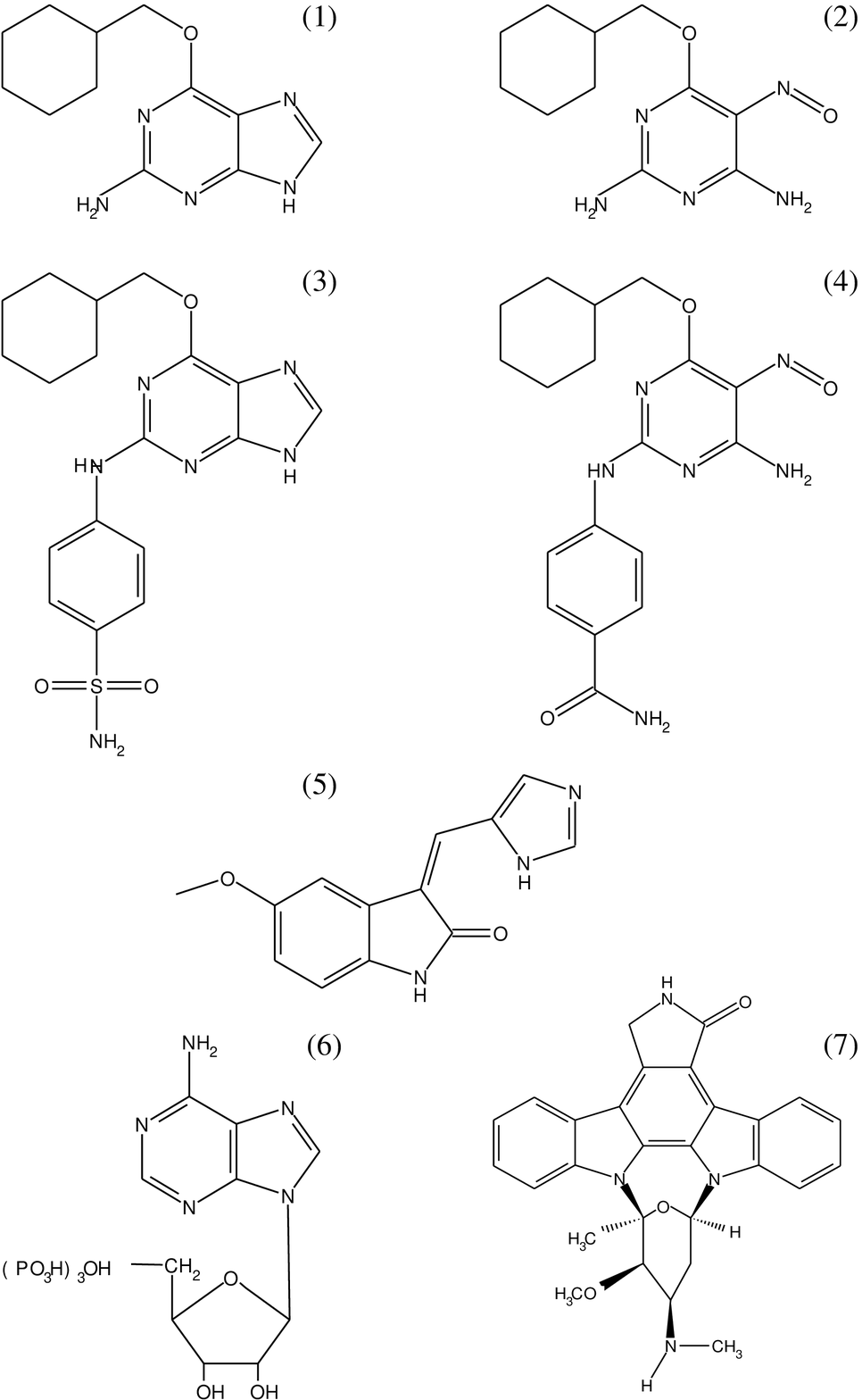}
\vfill  L. Heady {\em et al.}, Figure 1
\end{center}

\clearpage
\newpage

\begin{center}
\includegraphics[width=15.0cm]{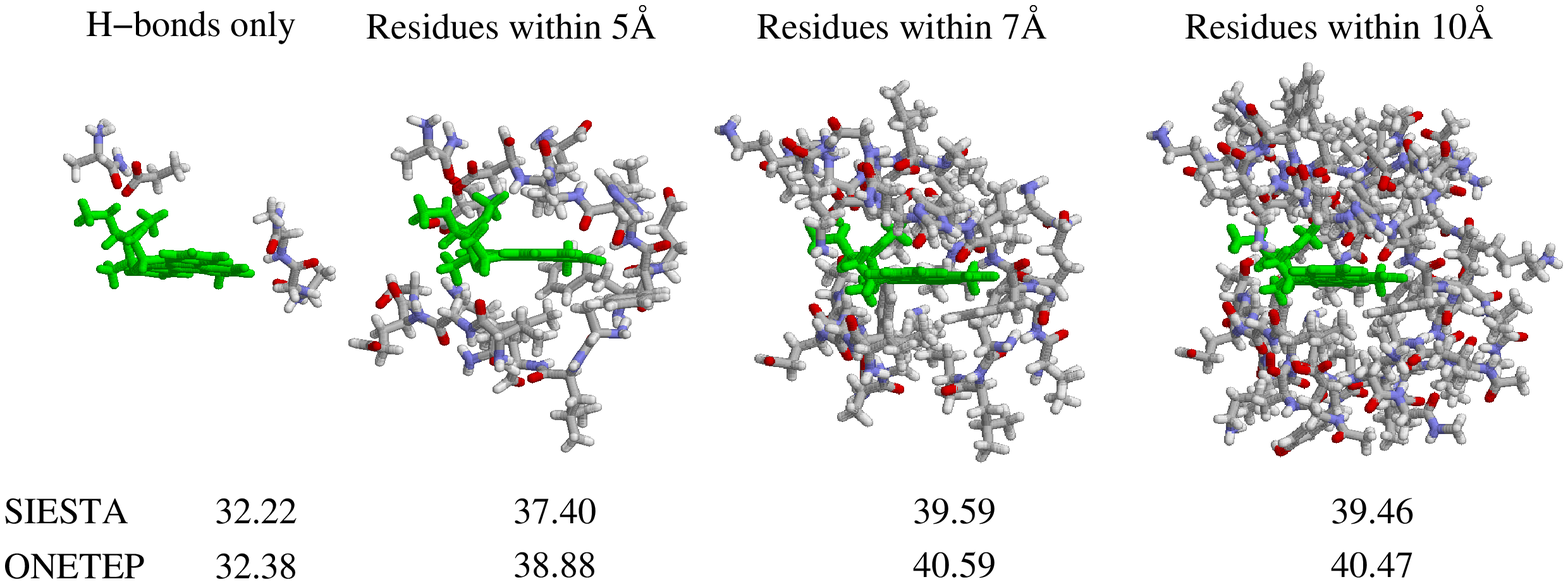}
\vfill  L. Heady {\em et al.}, Figure 2
\end{center}

\clearpage

\begin{center}
\includegraphics[width=12.0cm]{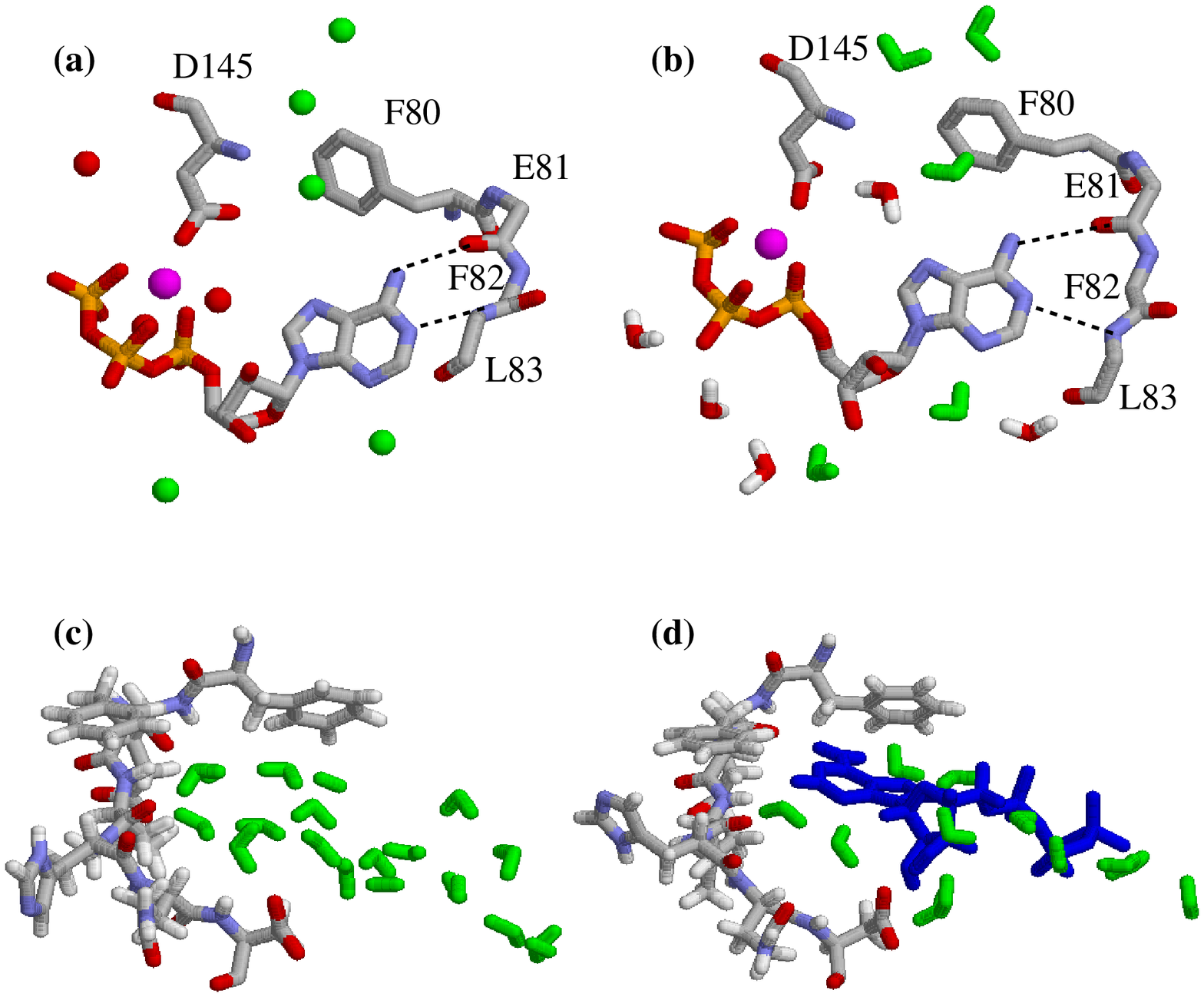}
\vfill  L. Heady {\em et al.}, Figure 3
\end{center}

\clearpage

\begin{center}
\includegraphics[width=8.0cm]{fig4.eps}
\vfill  L. Heady {\em et al.}, Figure 4
\end{center}

\clearpage

\begin{center}
\includegraphics[width=15.0cm]{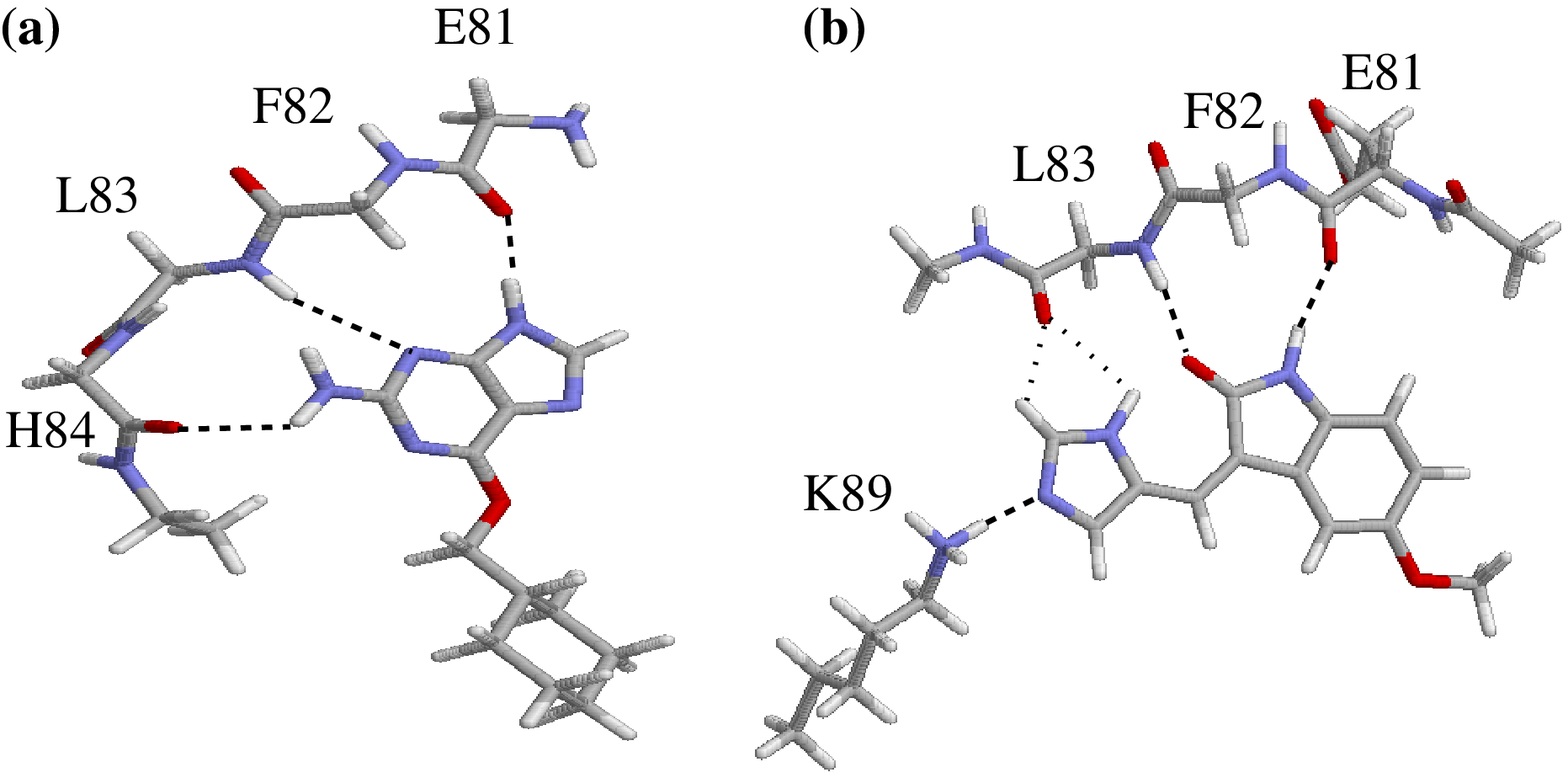}
\vfill  L. Heady {\em et al.}, Figure 5
\end{center}

\clearpage

\begin{center}
\hspace*{-1cm}
\includegraphics[width=15.0cm]{fig6.eps}
\vfill  L. Heady {\em et al.}, Figure 6
\end{center}

\clearpage

\begin{center}
\includegraphics[width=12.0cm]{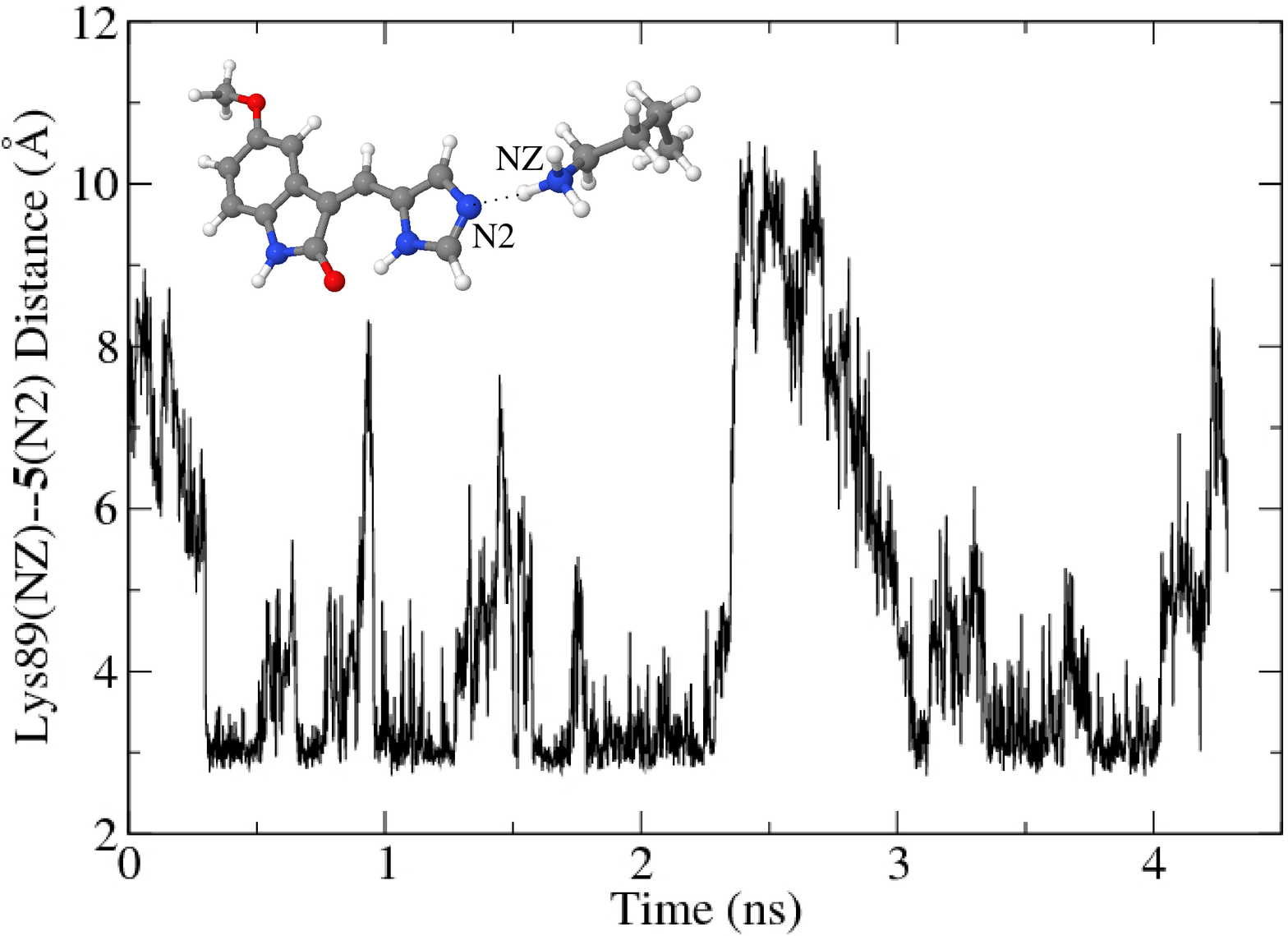}
\vfill  L. Heady {\em et al.}, Figure 7
\end{center}

\clearpage

\begin{center}
\includegraphics[width=8cm]{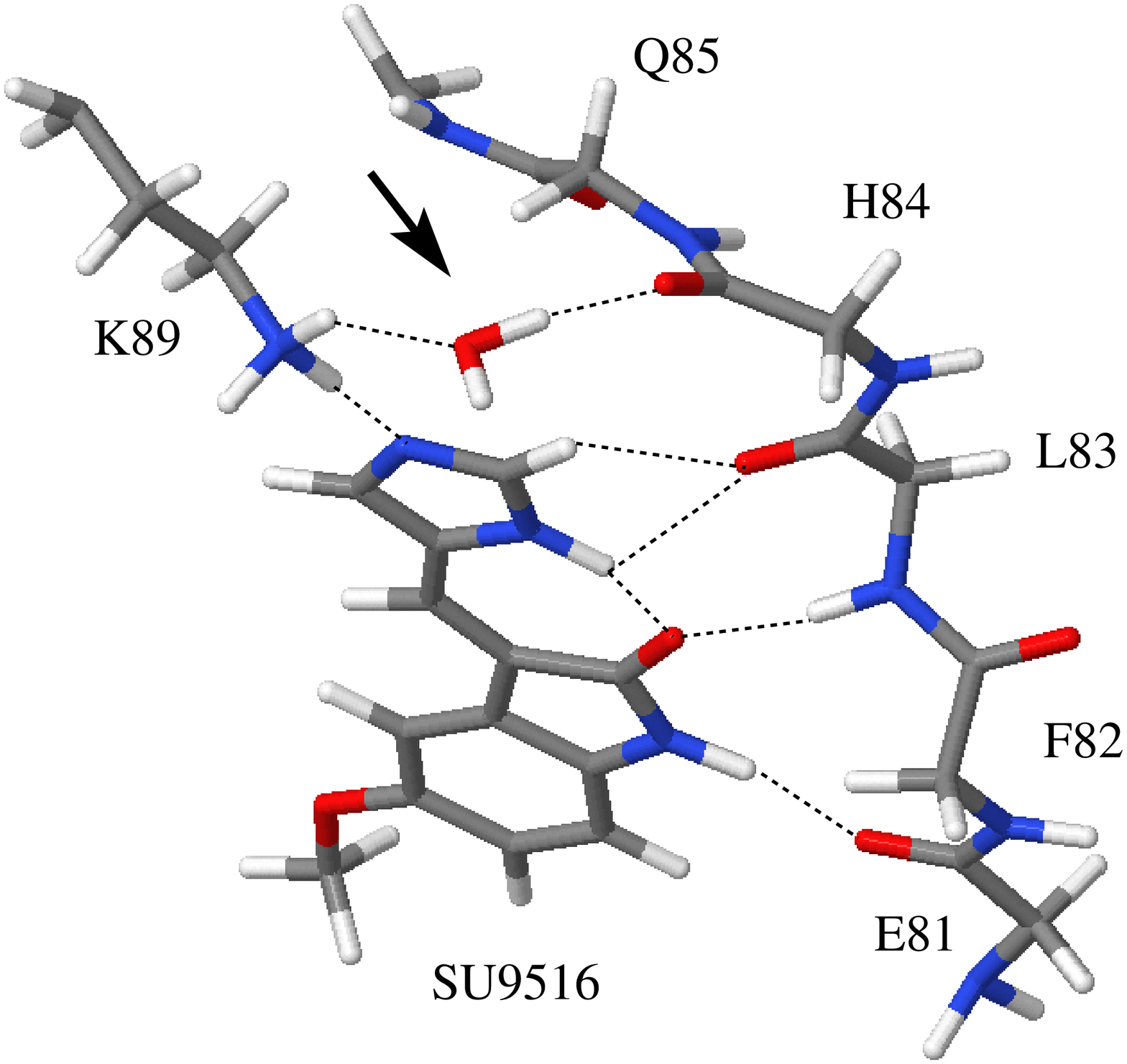}
\vfill  L. Heady {\em et al.}, Figure 8
\end{center}

\cleardoublepage
\newpage
\thispagestyle{empty}
\begin{center}

{\large \bf Table of Contents Graphic}

\vspace{1cm}

\includegraphics{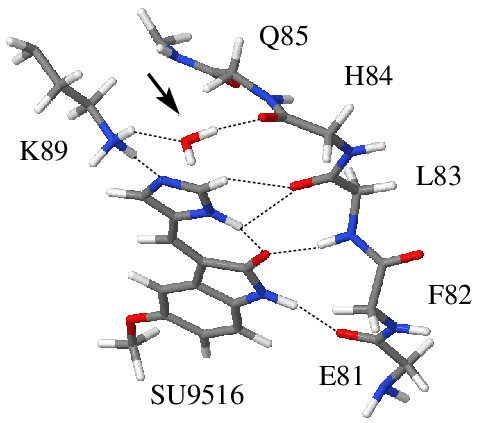}

\end{center}


\begin{thebibliography}{10}

\bibitem{Manning_2002}
Manning, G.; Whyte, D. B.; Martinez, R.; Hunter, T.; Sudarsanam, S.
The protein kinase complement of the human genome
{\em Science} {\bf 2002}, {\em 298}, 1912--1934.

\bibitem{Morgan_95}
 Morgan, D. O.
 Principles of CDK regulation.
 {\em Nature} {\bf 1995}, {\em 374}, 131--134.

\bibitem{Loog_05}
 Loog, M.; Morgan, D. O.
 Cyclin specificity in the phosphorylation of cyclin-dependent
 kinase substrates.
 {\em Nature} {\bf 2005}, {\em 434}, 104--108.

\bibitem{Brown_99}
 Brown, N. R.; Noble, M. E. M.; Lawrie, A. M.; Morris, M. C.;
 Tunnah, P.; Divita, G.; Johnson, L. N.; Endicott, J. A.
 Effects of Phosphorylation of Threonine 160 on Cyclin-dependent
 Kinase 2 Structure and Activity.
 {\em J. Biol. Chem.} {\bf 1999}, {\em 274}, 8746--8756.

\bibitem{Russo_96}
 Russo, A. A.; Jeffrey, P. D.; Pavletich, N. P.
 Structural basis of cyclin-dependent kinase activation by phosphorylation.
 {\em Nat. Struct. Biol.} {\bf 1996}, {\em 3}, 696--700.

\bibitem{Vermeulen_03}
 Vermeulen, K.; Van Bockstaele, D. R.; Berneman, Z. N.
 The cell cycle: a review of regulation, deregulation
 and therapeutic targets in cancer.
 {\em Cell Proliferat.} {\bf 2003}, {\em 36}, 131--149.

\bibitem{Huwe_03}
 Huwe, A.; Mazitschek, R.; Giannis, A.
 Small molecules as Inhibitors of Cyclin-Dependent Kinases
 {\em Angew. Chem. Int. Ed.} {\bf 2003}, {\em 42}, 2122--2138.

\bibitem{Wadler_01}
 Wadler, S.
 Perspectives for cancer therapies with CDK2 inhibitors.
 {\em Drug Resist. Updates} {\bf 2001} {\em 4}, 347--367.

\bibitem{Noble_04}
 Noble, M. E. M.; Endicott, J. A.; Johnson, L. N.
 Protein Kinase Inhibitors: Insights into Drug Design from Structure.
 {\em Science} {\bf 2004}, {\em 303}, 1800--1805.

\bibitem{Bain_03}
 Bain, J.; McLauchlan, H.; Elliott, M.; Cohen, P.
 The specificities of protein kinase inhibitors: an update.
 {\em Biochem. J.} {\bf 2003} {\em 371}, 199--204.

\bibitem{Park_04}
 Park, H.; Yeom, M. S.; Lee, S.
 Loop Flexibility and Solvent Dynamics as Determinants for the Selective
 Inhibition of Cyclin-Dependent Kinase 4: Comparative Molecular Dynamics
 Simulation Studies of CDK2 and CDK4.
 {\em ChemBioChem} {\bf 2004}, {\em 5}, 1662--1672.

\bibitem{Ikuta_01}
 Ikuta, M.; Kamata, K.; Fukasawa, K.; Honma, T.; Machida, T.; Hirai, H.;
 Suzuki-Takahashi, I.; Hayama, T.; Nishimura, S.
 Crystallographic Approach to Identification of Cyclin-Dependent
 Kinase 4 (CDK4)-specific Inhibitors by Using CDK4 Mimic CDK2 Protein.
 {\em J. Biol. Chem.} {\bf 2001} {\em 276}, 27548--27554.

\bibitem{Soler_02}
 Soler, J. M.; Artacho, E.; Gale, J. D.; Garc\'{i}a, A.; Junquera, J.;
 Ordej\'{o}n, P.; S\'{a}nchez-Portal, D.
 The \textsc{Siesta} method for {\em ab initio} order-n materials simulation.
 {\em J. Phys.-Cond. Mat.} {\bf 2002} {\em 14}, 2745--2779.

\bibitem{Skylaris_05}
 Skylaris, C.-K.; Haynes, P. D.; Mostofi, A. A.; Payne, M. C.
 Introducing \textsc{onetep}: Linear-scaling density functional
 simulations on parallel computers.
 {\em J. Chem. Phys.} {\bf 2005} {\em 122}, 084119.

\bibitem{Hardcastle_04}
 Hardcastle, I. R.; Arris, C. E.; Bentley, J.; Boyle, F. T.; Chen, Y. H.;
Curtin, N. J.;
 Endicott, J. A.; Gibson, A. E.; Golding, B. T.; Griffin, R. J.; Jewsbury, P.;
 Menyerol, J.; Mesguiche, V.; Newell, D. R.; Noble, M. E. M.; Pratt, D. J.;
 Wang, L. Z.; Whitfield, H. J.
 N$^2$-Substituted O$^6$-Cyclohexylmethylguanine Derivatives:
 Potent inhibitors of Cyclin-Dependent kinases 1 and 2
 {\em J. Med. Chem.} {\bf 2004}, {\em 47}, 3710--3722.

\bibitem{Gibson_02}
 Gibson, A. E.; Arris, C. E.; Bentley, J.; Boyle, F. T.; Curtin, N. J.;
 Davies, T. G.; Endicott, J. A.; Golding, B. T.; Grant, S.; Griffin, R. J.;
 Jewsbury, P.; Johnson, L. N.; Mesguiche, V.; Newell, D. R.;
 Noble, M. E. M.; Tucker, J. A.; Whitfield, H. J.
 Probing the ATP Ribose-Binding domain of Cyclin-Dependent Kinases
 1 and 2 with $O^6$-Substituted Guanine Derivatives.
 {\em J. Med. Chem.} {\bf 2002} {\em 45}, 3381--3393.

\bibitem{Sayle_03}
 Sayle, K. L.; Bentley, J.; Boyle, F. T.; Calvert, A. H.; Cheng, Y. Z.;
 Curtin, N. J.; Endicott, J. A.; Golding, B. T.; Hardcastle, I. R.;
 Jewsbury, P.; Mesguiche, V.; Newell, D. R.; Noble, M. E. M.; Parsons, R. J.;
 Pratt, D. J.; Wang, L. Z.; Griffin, R. J.
 Structure-based design of 2-arylamino-4-cyclohexylmethyl-5-nitroso-%
 6-aminopyrimidine inhibitors of cyclin-dependent kinases 1 and 2.
 {\em Bioorg. Med. Chem. Lett.} {\bf 2003}, {\em 13}, 3079--3082.

\bibitem{Legraverend_00}
 Legraverend, M.; Tunnah, P.; Noble, M.; Ducrot, P.; Ludwig, O.;
 Grierson, D.S.; Leost, M.; Meijer, L.; Endicott, J.
 Cyclin-Dependent Kinase Inhibition by New C-2 Alkynylated Purine
 Derivatives and Molecular Structure of a CDK2-Inhibitor Complex.
 {\em J. Med. Chem.} {\bf 2000} {\em 43}, 1282--1292.

\bibitem{Gray_98}
 Gray, N.S.; Wodicka, L.; Thunnissen, A.-M. W. H.; Norman, T. C.;
 Kwon, S. J.; Espinoza, F. H.; Morgan, D. O.; Barnes, G.; LeClerc, S.;
 Meijer, L.; Kim, S. H.; Lockhart, D. J.; Schultz, P. G.
 Exploiting chemical libraries, structure, and genomics in the search
 for kinase inhibitors.
 {\em Science} {\bf 1998} {\em 281}, 533--538.

\bibitem{Otyepka_00}
 Otyepka, M.; Krystof, V.; Havlicek, L.; Siglerova, V.; Strnad M.; Koca, J.
 Docking-Based development of Purine-like Inhibitors of
 Cyclin-Dependent Kinase 2.
 {\em J. Med. Chem.} {\bf 2000} {\em 43}, 2506--2513.

\bibitem{Gabb_97}
 Gabb, H. A.; Jackson, R. M.; Sternberg,  M. J. E.
 Modelling Protein Docking using Shape Complementarity,
 Electrostatics and Biochemical Information.
 {\em J. Mol. Biol.} {\bf 1997} {\em 272}, 106--120.

\bibitem{Ducrot_00}
 Ducrot, P.; Legraverend, M.; Grierson, D. S.
 3D-QSAR CoMFA on Cyclin-Dependent Kinase Inhibitors.
 {\em J. Med. Chem.} {\bf 2000} {\em 43}, 4098--4108.

\bibitem{Mancera_04}
 Mancera, R. L.; K\"allblad, P.; Todorov, N. P.
 Ligand-protein docking using a quantum stochastic tunneling optimization
method.
 {\em J. Comput. Chem.} {\bf 2004}, {\em 25}, 858--864.

\bibitem{Kriz_04}
 Kriz, Z.; Otyepka, M.; Bartova, I.; Koca J.
 Analysis of CDK2 Active-Site Hydration: A Method to Design New Inhibitors.
 {\em Proteins} {\bf 2004} {\em 55}, 258--274.

\bibitem{Bartova_04}
 Bartova, I.; Otyepka, M.; Kriz, Z.; Koca, J.
 Activation and inhibition of cyclin-dependent kinase-2 by phosphorylation;
 a molecular dynamics study reveals the functional importance of the
 glycine-rich loop.
 {\em Protein Sci.} {\bf 2004} {\em 6}, 1449--1457.

\bibitem{Otyepka_02}
 Otyepka, M.; Kriz, Z.; Koca, J.
 Dynamics and Binding Modes of Free CDK2 and its Two Complexes
 with Inhibitors Studied by Computer Simulations.
 {\em J. Biomol. Struct. Dyn.} {\bf 2002} {\em 20}, 141--154.

\bibitem{Cavalli_01}
 Cavalli, A.; Dezi, C.; Folkers, G.; Scapozza, L.; Recanatini, M.
 Three-Dimensional Model of the Cyclin-Dependent Kinase 1 (CDK1): Ab initio
 Active Site Parameters for Molecular Dynamics Studies of CDKs
 {\em Proteins} {\bf 2001}, {\em 45}, 478--485.

\bibitem{Ireta_04}
 Ireta, J.; Neugebauer, J.; Scheffler M.
 On the accuracy of DFT for describing hydrogen bonds: dependence on
 the bond directionality.
 {\em J. Phys. Chem. A} {\bf 2004}, {\em 108}, 5692--5698.

\bibitem{Arris_00}
 Arris, C. E.; Boyle, F. T.; Calvert, A. H.; Curtin, N. J.; Endicott, J. A.;
 Garman, E. F.; Gibson, A. E.; Golding, B. T.; Grant, S.; Griffin, R. J.;
 Jewsbury, P.; Johnson, L. N.; Lawrie, A. M.; Newell, D. R.; Noble, M. E. M.;
 Sausville, E. A.; Schultz, R.; Yu, W.
 Identification of Novel Purine and Pyrimidine Cyclin-Dependent Kinase
 Inhibitors with Distinct Molecular Interactions and Tumor Cell Growth
 Inhibition Profiles.
 {\em J. Med. Chem.} {\bf 2000}, {\em 43}, 2797--2804.

\bibitem{Davies_02}
 Davies, T. G.; Bentley, J.; Arris, C. E.; Boyle, F. T.; Curtin, N. J.;
 Endicott, J. A.; Gibson, A. E.; Golding, B. T.; Griffin, R. J.;
 Hardcastle, I. R.; Jewsbury, P.; Johnson, L. N.; Mesguiche, V.;
 Newell, D. R.; Noble, M. E. M.; Tucker, J. A.; Wang, L.; Whitfield, H. J.
 Structure-based Design of a potent purine-based cyclin-dependent
 kinase inhibitor.
 {\em Nat. Struct. Biol.} {\bf 2002},{\em 9}, 745--749.

\bibitem{Lane_01}
 Lane, M. E.; Yu, B.; Rice, A.; Lipson, K. E.; Liang, C.; Sun, L.;
 Tang, C.; McMahon, G.; Pestell, R. G.; Wadler, S.
 A novel CDK2-selective inhibitor, SU9516, induces apoptosis in
 colon carcinoma cells
 {\em Cancer Res.} {\bf 2001}, {\em 61}, 6170--6177.

\bibitem{Moshinsky_03}
 Moshinsky, D. J.; Bellamacina, C. R.; Boisvert, D. C.; Huang, P.;
 Hui, T.; Jancarik, J.; Kim, S.-H.; Rice, A. G.
 SU9516: Biochemical Analysis of CDK Inhibition and Crystal Ctructure
 in Complex with CDK2.
 {\em Biochem. Bioph. Res. Co.} {\bf 2003}, {\em 310}, 1026--1031.

\bibitem{Cheng_73}
 Cheng, Y.-C.; Prusoff, W. H.
 Relationship between the inhibition constant K$_{i}$ and the
concentration of inhibitor which causes 50 per cent inhibition
(IC$_{50}$) of an enzymatic reaction.
{\em Biochem. Parmacol.} {\bf 1973}, {\em 22}, 3099--3108.

\bibitem{Wang_01}
 Wang, J.; Morin, P.; Wang, W.; Kollman, P. A.
 Use of MM-PBSA in Reproducing the Binding Free Energies to HIV-1 RT of TIBO 
 Derivatives and Predicting the Binding Mode to HIV-1 RT of Efavirenz by Docking 
 and MM-PBSA
 {\em J. Am. Chem. Soc.} {\bf 2001}, {\em 123}, 5221--5230.

\bibitem{Cramer_99}
Cramer C. J.; Truhlar, D. J.
Implicit Solvation Models: Equilibria, Structure, Spectra, and Dynamics
{\em Chem. Rev.} {\bf 1999}, {\bf 99}, 2161--2200.

\bibitem{Jayaram_98}
 Jayaram, B.; Sprous, D.; Beveridge, D. L.
 Solvation Free Energy of Biomacromolecules: Parameters for a Modified
 Generalized Born Model Consistent with the Amber Force Field
 {\em J. Phys. Chem. B} {\bf 1998}, {\em 102}, 9571--9576.

\bibitem{Sitkoff_94}
 Sitkoff, D.; Sharp, K. A.; Honig, B.
 Accurate Calculation of Hydration Free Energies Using Macroscopic Solvent
Models.
 {\em J. Phy. Chem.} {\bf 1994}, {\em 98}, 1978--1988.

\bibitem{Fernandez_04}
 Fen\'andez-Serra, M. V.; Artacho, E.
 Network equilibration and first principles liquid water.
 {\em J. Chem. Phys.} {\bf 2004} {\em 121}, 11136--11144.

\bibitem{Galli_04}
 Schwegler, E.; Grossman, J.C.; Gygi, F.; Galli, G.
 Towards an assessment of the accuracy of density functional theory 
 for first principles simulations of water. II
 {\em J. Chem. Phys.} {\bf 2004} {\em 121}, 5400--5409.
%
 Grossman, J.C.; Schwegler, E.; Draeger, E.W.; Gygi, F.; Galli, G.
 Towards an assessment of the accuracy of density functional theory 
 for first principles simulations of water. I
 {\em J. Chem. Phys.} {\bf 2004} {\em 120}, 300--311.

\bibitem{Vondele_05}
VandeVondele, J.; Mohamed, F.; Krack, M.; Hutter, J.; Sprik, M.;
Parrinello, M. 
The influence of temperature and density functional models in ab initio
molecular dynamics simulation of liquid water
 {\em J. Chem. Phys.} {\bf 2005}, {\em 122}, 014515.
 
\bibitem{Mancera_wat}
 Mancera, R. L.
 A new explicit hydration penalty score for ligand-protein interactions
 {\em Chem. Phys. Lett.} {\bf 2004}, {\em 399}, 271--275.

\bibitem{Lawrie_97}
 Lawrie, A. M.; Noble, M. E. M.; Tunnah, P.; Brown, N. R.;
 Johnson, L. N.; Endicott, J. A.
 Protein kinase inhibition by staurosporine revealed in details
 of the molecular interaction with CDK2.
 {\em Nat. Struct. Biol.} {\bf 1997}, {\em 4}, 796--801.

 \bibitem{Schulze_96}
 Schulze-Gahmen, U.; De~Bondt H. L.; Kim, S.-H.
 High-Resolution Crystal Structures of Human Cyclin-Dependent Kinase 2
 with and without ATP: Bound Waters and Natural Ligand as Guides for
 Inhibitor Design.
 {\em J. Med. Chem.} {\bf 1996}, {\em 39}, 4540--4546.

\bibitem{Phe_82}
In particular, a steric clash between all inhibitors and the
Phe~82 residue is evident after including the small model
minimised at the DFT level in the larger pocket model
obtained after force-field minimisation from a MD snapshot.
%
This is due to the much shorter van der Waals distance between
hydrophobic residues obtained with classical potentials
with respect to the corresponding distances obtained at
the GGA-DFT level.
%
For this reason, since the contributions of direct van der Waals
interactions with the Phe~82 residue are expected to be about
the same for all inhibitors, this residue is substituted with an
alanine residue in all binding energy calculations presented.

\bibitem{Polarization}
 (a)  Greatbanks, S.P.; Gready, J.E.; Limaye, A.C.; Rendell, A.P.
 Comparison of enzyme polarization of ligands and charge-transfer 
 effects for dihydrofolate reductase using point-charge embedded 
 ab initio quantum mechanical and linear-scaling semiempirical 
 quantum mechanical methods.
 {\em J. Comput. Chem.} {\bf 2000}, {\em 21}, 788--811. 
%
 (b) Muzet, N.; Guillot, B.; Jelsch, C.; Howard, E.; Lecomte, C.
 Electrostatic complementarity in an aldose reductase complex 
 from ultra-high-resolution crystallography and first-principles 
 calculations.
 {\em Proc. Natl. Acad. Sci. USA} {\bf 2003}, {\em 100}, 8742--8747.
%
 (c) Sulpizi, M.; Laio, A.; VandeVondele, J.; Cattaneo, A.; 
  Rothlisberger, U.; Carloni, P. 
  Reaction mechanism of caspases: Insights from QM/MM Car-Parrinello simulations 
  {\em Proteins} {\bf 2003}, {\em 52}, 212--224.


\bibitem{Boltz_fac}
We note that an accurate determination of the time fractions
of different binding modes would require the calculation of 
their associated Boltzmann factors, which is computationally
feasible only at the classical level.
%
However, the accuracy of this approach relies on the precision of
the calculated energy values of different binding configurations.
%
These, in turn, are likely to be affected by subtle electronic 
effects which are difficult to describe using a fixed-parameter 
force field approach.
%
Therefore, our choice here is to use the force field to obtain
an estimate of the binding mode time fractions directly from
MD simulations, while all energy values are computed at the
DFT level.

\bibitem{Hobza_05}
Daabkowska, I.; Jurecka, P.; Hobza, P.
On geometries of stacked and H-bonded nucleic acid base pairs determined 
at various DFT, MP2, and CCSD(T) levels up to the CCSD(T)/complete basis 
set limit level.
{\em J. Chem. Phys.} {\bf 2005}, {\em 122}, 204322.

\bibitem{Elstner_01}
Elstner, M.; Hobza, P.; Frauenheim T.; Suhai, S.; Kaxiras, E.
Hydrogen-bonding and stacking interactions of nucleic acid base
pairs: A density-functional-theory based treatment.
{\em J. Chem. Phys.} {\bf 2001}, {\em 114}, 5149--5155. 

\bibitem{Barrat_05}
Barratt, E.; Bingham, R. J.; Warner, D. J.; Laughton, C. A.;
Phillips, S. E. V.; Homans, S: W.
Van der Waals interactions dominate ligand-protein association
in a protein binding site occluded from solvent water.
{\em J. Am. Chem. Soc.} {\bf 2005}, {\em 127}, 11827-11834. 

\bibitem{Honma_01}
Honma, T.; Yoshizumi, T.; Hashimoto, N.; Hayashi, K.; Kawanishi, N.;
Fukasawa, K.; Takaki, T.; Ikeura, C.; Ikuta, M.; Suzuki-Takahashi, I.;
Hayama, T.; Nishimura, S.; Morishima, H.
A Novel Approach for the Development of Selective CDK4 Inhibitors:
Library Design Based on Locations of CDK4 Specific Amino Acid Residues
{\em  J. Med. Chem.} {\bf 2001}, {\em 44}, 4628--4640.

\bibitem{Hamdouchi_05}
Hamdouchi, C.; Zhong, B.; Mendoza, J.; Collins, E.; Jaramillo, C.;
De Diego, J. E.; Robertson, D.; Spencer, C. D. ; Anderson, B. D.; 
Watkins, S. A.; Zhang, F.; Brooks, H. B.
Structure-based design of a new class of highly selective 
aminoimidazo[1,2-$\alpha$]pyridine-based inhibitors of 
cyclin dependent kinases
{\em Bioorg. Med. Chem. Lett.} {\bf 2005}, {\em 15},1943--1947.

\bibitem{Mancera_06}
Kelly, M. D.; Mancera, R. L. 
Comparative analysis of the surface interaction properties
of the binding sites of CDK2, CDK4 and ERK2.
{\em ChemMedChem} {\bf 2006}, {\em 1}, 366--375.

\bibitem{Payne_92}
 Payne, M. C.; Teter, M. P.; Allan, D. C.; Arias, T. A.; Joannopoulos, J. D.
 Iterative minimisation techniques for ab initio total-energy
 calculations: molecular dynamics and conjugate gradients.
 {\em Rev. Mod. Phys.} {\bf 1992}, {\em 64}, 1045--1097.

\bibitem{Perdew_96}
 Perdew, J. P.; Burke, K.; Ernzerhof, M.
 Generalized gradient approximation made simple.
 {\em Phys. Rev. Lett.} {\bf 1996}, {\em 77}, 3865--3868.

\bibitem{Segall_02}
 Segall M. D.; Lindan, P. J. D.; Probert, M. J.; Pickard, C. J.;
 Hasnip, P. J.; Clark, S. J.; Payne, M. C.
 First-principles simulation: ideas, illustrations and the \textsc{castep} code.
 {\em J. Phys.-Condens. Mat.} {\bf 2002}, {\em 14}, 2717--2744.

\bibitem{Vanderbilt_90}
 Vanderbilt, D.
 Soft self-consistent pseudopotential in a generalized eigenvalue formalism.
 {\em Phys. Rev. B} {\bf 1990}, {\em 41}, 7892--7895.

\bibitem{Hamann_79}
 Hamann, D. R.; Schl\"{u}ter, M.; Chiang, C.
 Norm-conserving pseudopotentials.
 {\em Phys. Rev. Lett.} {\bf 1979}, {\em 43}, 1494--1497.

\bibitem{Troullier_91}
 Troullier, N.; Martins J. L.
 Efficient pseudopotentials for plane-wave calculations.
 {\em Phys. Rev. B} {\bf 1991}, {\em 43}, 1993--2006.

\bibitem{Anglada_02}
 Anglada, E.; Soler, J. M.; Junquera, J.; Artacho, E.
 Systematic generation of finite-range atomic basis set
 for linear-scaling calculations. 
 {\em Phys. Rev. B} {\bf 2002} {\em 66}, 205101.

\bibitem{Boys_70}
 Boys, S. F.; Bernardi, F.
 The calculation of small molecular interactions by the differences of
 separate total energies. Some procedures with reduced errors.
 {\em Mol. Phys.} {\bf 1970} {\em 19}, 553--566.

\bibitem{Skylaris_02-1}
 Skylaris, C.-K.; Mostofi, A. A.; Haynes, P. D.; Di\'eguez, O.; Payne, M. C.
 Nonorthogonal generalised Wannier function pseudopotential plane-wave method.
 {\em Phys. Rev. B} {\bf 2002}, {\em 66}, 035119.

\bibitem{Mostofi_03}
 Mostofi, A. A.; Haynes, P. D.; Skylaris, C.-K.; Payne, M. C.
 Preconditioned iterative minimisation for linear-scaling electronic
 structure calculations.
 {\em J. Chem. Phys.} {\bf 2003}, {\em 119}, 8842--8848.
 
\bibitem{Haynes_05}
 Skylaris, C.-K.; Haynes, P. D.; Mostofi, A. A.; Payne, M. C.
 Using \textsc{Onetep} for accurate and efficient O(N) density
functional calculations.
{\em J. Phys.: Condens. Matter} {\bf 2005}, {\em 17}, 5757--5769. 

\bibitem{amber7}
 Case, D. A.; Pearlman, D. A.; Caldwell, J. W.; Cheatham, T. E.; Wang, J.;
 Ross, W. S.; Simmerling, C. L.; Darden, T. A.; Merz, K. M.; Stanton, R. V.;
 Cheng, A. L.; Vincent, J. J.; Crowley, M.; Tsui, V.; Gohlke, H.;
 Radmer, R. J.; Duan, Y.; Pitera, J.; Massova, I.; Seibel, G. L.;
 Singh, U. C.; Weiner, P. K.; Kollman, P. A.
 AMBER 7.
 {\em University of California, San Francisco,} {\bf 2002}.

\bibitem{Wang_00}
 Wang, J. M.; Cieplak, P.; Kollman, P. A.
 How well does a restrained electrostatic potential (RESP) model perform in
 calculating conformational energies of organic and biological molecules?
 {\em J. Comput. Chem.} {\bf 2000}, {\em 21}, 1049--1074.

\bibitem{Cornell_95}
 Cornell, W. D.; Cieplak, P.; Bayly, C. I.; Gould, I. R.;  Merz, K. M.;
 Ferguson, D. M.; Spellmayer, D. C.; Fox, T.; Caldwell, J. W.; Kollman, P. A.
 A second generation force-field for the simulation of proteins,
 nucleic acids, and organic molecules
 {\em J. Am. Chem. Soc.} {\bf 1995} {\em 117}, 5179--5197.

\bibitem{Todorov_03}
 Todorov, N. P.; Mancera, R. L.; Monthoux, P. H.
 A new quantum stochastic tunnelling optimisation method
 for protein-ligand docking.
 {\em Chem. Phys. Lett.} {\bf 2003}, {\em 369}, 257--263.

\bibitem{Gehlhaar_95}
 Gehlhaar, D. K.; Verkhivker, G. M.; Rejto, P. A.; Sherman, C. J.;
 Fogel, D. B.; Fogel, L. J.; Freer, S. T.
 Molecular Recognition of the Inhibitor AG-1343 by HIV-1 Protease --
 Conformationally Flexible Docking by Evolutionary Programming.
 {\em Chem. Biol.} {\bf 1995}, {\em 2}, 317--324.

\bibitem{Boehm_94}
 B\"ohm, H.-J.
 The Development of a Simple Empirical Scoring Function to Estimate
 the Binding Constant for a Protein Ligand Complex of Known 3-Dimensional
 Structure
 {\em J. Comput. Aid. Mol. Des.}{\bf 1994 } {\em 8}, 243--256.

\bibitem{Stahl_01}
 Stahl, M.; Rarey, M.
 Detailed analysis of scoring functions for virtual screening
 {\em J. Med. Chem.} {\bf 2001}, {\em 44}, 1035--1042.


\end{thebibliography}
\end{document}